%% file: lat16-proceedings.tex
\documentclass[nohyper]{PoS}
\pdfoutput=1

\usepackage{amsmath,amsfonts,amssymb,amsbsy}
\usepackage{booktabs}
\usepackage[margin=4pt]{subcaption}

\usepackage{macros_alpha,defs}

\title{The $\Lambda$-parameter in 3-flavour QCD and $\alpha_s(m_Z)$ by the ALPHA collaboration}

\ShortTitle{$\Lambda^{(3)}_{\overline{\rm MS}}$ and $\alpha_s(m_Z)$ by the ALPHA collaboration 
$\hphantom{0123456789012345678901234}\mathrm{S.\,Sint,\, A.\,Ramos,\, R.\,Sommer}$}

\author{%
M.~Bruno$^{\,a}$, %
M.~Dalla~Brida$^{\,b,c}$, %
P.~Fritzsch$^{\,d,e}$, %
T.~Korzec$^{\,f}$, %
{A.~Ramos}$^{\,d\,\ast}$, %
S.~Schaefer$^{\,c}$, %
H.~Simma$^{\,c}$, %
{S.~Sint}$^{\,g\,\ast}$ and %
{R.~Sommer}$^{\,c,h\,}$\thanks{Speakers}
\vskip0.25em\\
\llap{$^a$}Physics Department, Brookhaven National Laboratory, Upton, NY 11973, USA \\
\llap{$^b$}Dipartimento di Fisica, Universit\`a di Milano-Bicocca \&
INFN, sezione di Milano-Bicocca, 
Piazza della Scienza 3, I-20126 Milano, Italy\\
\llap{$^c$}NIC, DESY, Platanenallee~6, D-15738~Zeuthen, Germany\\
\llap{$^d$}PH-TH, CERN, CH-1211 Geneva, Switzerland\\
\llap{$^e$}Instituto de F\'{\i}sica Te{\'o}rica UAM/CSIC, Universidad Aut{\'o}noma de Madrid,\\
C/ Nicol{\'a}s Cabrera 13-15, Cantoblanco, Madrid 28049, Spain\\
\llap{$^f$}Department of Physics, Bergische Universit\"at Wuppertal, Gau\ss str. 20,\\
D-42119 Wuppertal, Germany\\
\llap{$^g$}School of Mathematics \& Hamilton Mathematics Institute, Trinity College, Dublin 2, Ireland\\
\llap{$^h$}Institut~f\"ur~Physik, Humboldt-Universit\"at~zu~Berlin, Newtonstr.~15, 12489~Berlin, Germany\\

\vskip0.25em\\
E-mail:~\email{mbruno@bnl.gov}, \hspace{1ex}
\email{mattia.dalla.brida@desy.de},\hspace{1ex}
\email{p.fritzsch@csic.es},\hspace{1ex}
\email{korzec@physik.hu-berlin.de},\hspace{1ex} 
\email{alberto.ramos@cern.ch},\hspace{1ex}
\email{stefan.schaefer@desy.de},\hspace{1ex}
\email{hubert.simma@desy.de},\hspace{1ex}
\email{sint@maths.tcd.ie},\hspace{1ex}
\email{rainer.sommer@desy.de} %
}

\abstract{We present results 
by the ALPHA collaboration for the $\Lambda$-parameter in 3-flavour QCD
and the strong coupling constant at the electroweak scale,
$\alpha_s(m_Z)$, in terms of hadronic quantities computed
on the CLS gauge configurations.
The first part of this proceedings contribution contains a review
of published material~\cite{Brida:2016flw,DallaBrida:2016kgh} and yields
the $\Lambda$-parameter in units of a low energy scale, $1/\lmax$. We then discuss
how to determine this scale in physical units from experimental data
for the pion and kaon decay constants. 
We obtain $\Lambda_\msbar^{(3)} = 332(14)\,\MeV$ which translates
to $\alpha_s(M_Z)=0.1179(10)(2)$ using perturbation theory
to match between 3-, 4- and 5-flavour QCD.
\vspace{2cm}
\begin{flushright}
CERN-TH-2016-262 \\ 
DESY 17-007 
\end{flushright}
}

\FullConference{34th annual International Symposium on Lattice Field Theory\\
		24-30 July 2016\\
		University of Southampton, UK}

\begin{document}

\input{talk_sint.tex}
\input{talk_ramos.tex}

\section{Hadronic scales}
\label{s:had}

It is left to fix $\Lhad$ in physical units from 
$\Lhad = (\Lhad m_\mathrm{had})^{(3)}/m_\mathrm{had}^\mathrm{exp}$
where $m_\mathrm{had}$ is an experimentally accessible low energy mass (scale)
and $(\Lhad m_\mathrm{had})^{(3)}$ is the dimensionless number computed
in QCD with three quark flavors. 
While it is most natural to use the proton
mass, $m_p$, it is not that easy to compute it with precision 
due to large statistical errors in the relevant correlation function 
at Euclidean times of 1~fm and larger and due to its complicated dependence
on the quark masses. Such technical limitations apply similarly 
to many other quantities. As explained in detail in
\cite{Sommer:2014mea} we are lead to choose the leptonic decay constant 
of pion and kaon for precision scale setting, 
even though their phenomenological values
$\fpi=130.4(2)\,\MeV$ and $\fK=156.2(7)\,\MeV$ depend on the knowledge
of $V_\mathrm{ud}$ and $V_\mathrm{us}$ \cite{Aoki:2016frl}.

In fact in order to express $\Lhad$ in physical units, we first 
relate $\fpi,\fK$ to an intermediate large volume scale $t_0^*$ and
then connect that to $\Lhad$.

\subsection{From $\pi$ and K decay constants to the reference scale $t_0^*$}

Our computation of hadronic scales is based on the
CLS large volume simulations with two degenerate light quarks, $m_u=m_d$, and one additional strange quark~\cite{Bruno:2014jqa}. 
In these simulations the trace, $m_u+m_d+m_s$, of the quark mass matrix $M$
is held constant~\cite{Bietenholz:2010jr} while varying $m_u=m_d$ in approaching the physical point defined by physical values for
$\mpi/\fpik,\,\mK/\fpik$. 

{\em Along this trajectory} in the quark mass plane 
the linear combination
\bes
  \fpik=(2\fK+\fpi)/3\,
\ees 
has a particularly simple dependence on the quark masses or equivalently on
their hadronic proxies
\begin{equation}
y_\pi = \frac{m_\pi^2}{(4 \pi \fpik)^2}\,,\quad
y_\mathrm{K} =\frac{m_\mathrm{K}^2}{(4 \pi \fpik)^2}\,. 
\end{equation} 
Namely, in the continuum limit, the expansion around the 
symmetric point, $y_\pi=y_\mathrm{K}=y_\mathrm{sym}$, reads
\bes
  \fpik=\fpik^\mathrm{symm}\,[\,(1 + h_2 (y_\pi-y_\mathrm{sym})^2 + 
   \rmO((y_\pi-y_\mathrm{sym})^3)\,]\, .
   \label{e:massdep1}
\ees 

Furthermore,  SU(3) chiral perturbation theory predicts 
the
quark mass dependence~\cite{Gasser:1984gg} free of low energy constants
in the form
\bes
  \fpik=\fpik^\mathrm{symm}\,[\,(1 +3L_\chi(y_\mathrm{sym}) - L_\chi(y_\pi) 
        - 2 L_\chi(y_\mathrm{K})]
       + \rmO(y^2) \,.
   \label{e:massdep2}
\ees 
Here the typical chiral logs 
\bes
L_\chi(y)  = y \, \log(y)\,, 
\ees 
appear.
Both forms \eq{e:massdep1} and \eq{e:massdep2} have been used 
for the extrapolation from the simulation points to the physical 
point~\cite{Bruno:2016plf}. They agree well within the statistical errors. 
Still, the small difference at the physical point is used as an
estimate of the  
remaining systematic error in the extrapolation.
\begin{figure}
  \centering
  \includegraphics[width=0.6\textwidth]{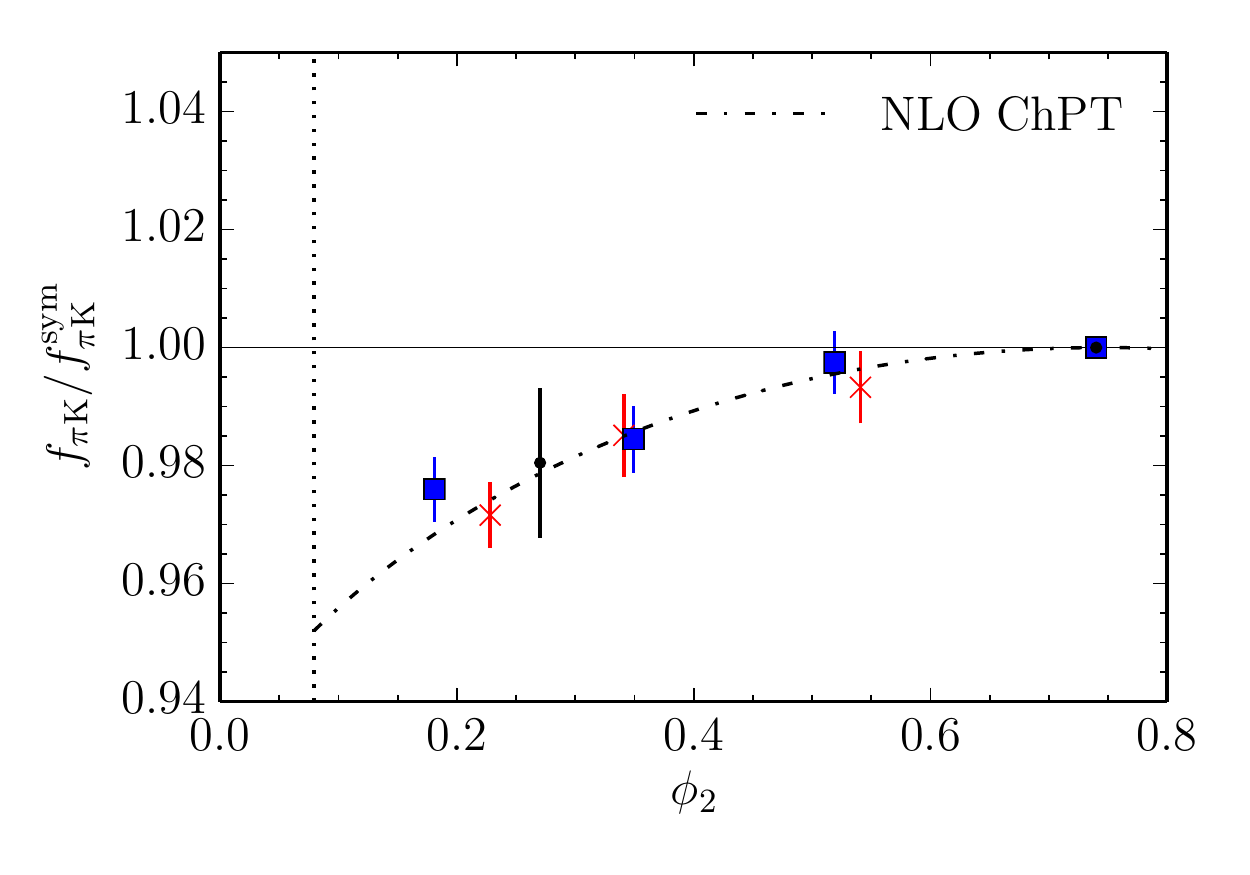}
  \caption{Chiral extrapolation of $f_{\pi K}$. In the horizontal axis
  we plot $\phi_2=8t_0m_\pi^2$, and we normalize the data with respect
  to the symmetric point. The data has been shifted to constant
  $\phi_4=1.11$.}
  \label{fig:chi}
\end{figure}

There are more details, e.g. the small cutoff effects have to be taken into 
account, cf. ref.~\cite{Bruno:2016plf}. There 
the physical $\fpik$ was then related to
$t_0^*$, the Gradient Flow scale, $t_0$ introduced by 
M.~L\"uscher \cite{Luscher:2010iy} at a particular 
mass point. This reference mass point is defined in terms
of the dimensionless variable
\bes
  \Phi_4= 8\,t_0 \,(\mK^2+\frac12\mpi^2)\,.  
\ees
We choose the symmetric line and set
\bes
  \left. \Phi_4\right|_{\mpi=\mpi^*} = 1.11 \text{\quad for\quad} m_u=m_d=m_s\,,
\ees
and
\bes
   t_0^* = 1.11 / [12 (\mpi^*)^2]\,.
\ees
Setting the scale with the phenomenological 
$\fpik$  yielded 
\bes
   (8t_0^*)^{1/2} = 0.413(5)({2}) \,\fm \,,
   \label{e:t0fm}
\ees
where the second error is the systematic error from the extrapolation to the 
physical point.
The scale $t_0^*$
is a good quantity to connect finite and large volumes 
and it is likely also a good one for other purposes.
Being defined in the mass-degenerate theory with
quark masses significantly heavier than the physical up and down quark
masses, there are only two parameters 
and, since $\mpi$ is around $400\,\MeV$, simulations
are easy and finite size effects are relatively small.

%%%%%%%%%%%%%%%%%%%%%%%%%%%%%%%%%%%%%%%%%%%%%%%%%%%%%%%%%%%%%%%%%%
\begin{table}
\begin{center}
\begin{tabular}{@{\extracolsep{0.2cm}}ccccccccc}
\toprule
$\beta$ &  $a\mq$ &  $\tilde\beta$ & $t_0^*/a^2$ & $\lmax/a$ &   $({t_0^*})^{-1/2} \lmax$ \\ 
\midrule 
    3.4000  &     0.0068  &     3.3985  &      2.862(  5) &  12.05(8) & 7.13(5)\\
    3.4600  &     0.0059  &     3.4587  &      3.662( 12) &  13.51(6) & 7.06(3)\\
    3.5500  &     0.0048  &     3.5490  &      5.166( 15) &  15.94(6) & 7.01(3)\\
    \it 3.5503  &     0       &     \it 3.5503  &                 &  16 & \\
    3.7000  &     0.0037  &     3.6992  &      8.596( 27) &  20.70(9) & 7.06(3)\\
    \it 3.7934  &     0       &     \it 3.7934  &                 &  24 & \\
    3.8500  &     0.0029  &     3.8494  &  {\it 13.880(220)} & 26.42(9) & 7.11(8)\\
    \it 3.9753  &     0       &     \it 3.9753  &                 &  32 & \\
\midrule 
%\multicolumn{3}{c}{sum:} & {\bf 5 M} & & {\bf 61 M}\\
\bottomrule
\end{tabular}
\end{center}
\caption{\label{t:t0} 
Results for $t_0^*/a^2$ of the large volume CLS runs at bare improved couplings
$\tilde\beta$. The  value of $t_0^*/a^2$ at $\beta=3.85$ 
is still very preliminary. Also numbers for $\lmax/a>13$ are still preliminary as
explained in \app{s:interpol}.
}
\end{table}
%%%%%%%%%%%%%%%%%%%%%%%%%%%%%%%%%%%%%%%%%%%%%%%%%%%%%%%%%%%%%%%%%%
%   3.400000  11.308074   0.099412   11.308074   0.099412  
%  beta     gsq for m(L/2)=0  err    gsq for m(L)=0     err 
%   3.550291  11.586074   0.083299      11.308074   0.083299 
%  beta     gsq for m(L/2)=0  err    gsq for m(L)=0     err 
%   3.793377  11.431629   0.062221      11.308074   0.062221 
%  beta     gsq for m(L/2)=0  err    gsq for m(L)=0     err 
%   3.975299  11.377574   0.058334      11.308074   0.058334 
   
% 12.0543   13.5074   15.9374   20.7032   26.4244
%  0.0754    0.0587    0.0561    0.0871    0.0926
%  7.1255    7.0585    7.0120    7.0614    7.1106
%  0.0452    0.0328    0.0267    0.0317    0.0763

These properties enable determinations of $t_0^*/a^2$ and $\Lhad/a$
in a large common range of lattice spacings $a$ and a subsequent 
controlled continuum extrapolation. We now describe this step in some detail.

\subsection{Three flavor $\Lambda$-parameter in physical units}

The large volume quantity $t_0^*$ is defined with finite (degenerate) quark masses.
In order to have $\rmO(a)$ improvement in its connection to the massless theory,
we need to combine $t_0^*/a^2$ and $\lmax/a$ at matching improved bare coupling 
\cite{Sint:1995ch,Luscher:1996sc},
\bes
     \tilde \beta &=& \beta_\mathrm{CLS} \,/\, (\,1+ \,a \tr M\,b_g^{(1)} \,2/\beta_\mathrm{CLS})\, +\, \rmO(1/\beta_\mathrm{CLS})\,,
     \label{e:betatilde}\\
     && \tr M = 3 \mq\,, \quad b_g^{(1)}  = 0.03600\,.
\ees
For the evaluation of the bare subtracted quark mass, $a\mq=1/(2\kappa) -1/(2\hopc))$,
we need the critical hopping parameter, $\hopc$. We estimate it by linear extrapolation in
$(a/L)^3$ of the critical hopping parameters defined and 
determined by setting the PCAC mass
on $(L/a)^4$ lattices to zero. This large $L$ extrapolation is carried out from the 
$\hopc$ for the two 
largest available lattices, namely $L/a=12,16$ \cite{Nf3tuning}. 
The relevant numbers are listed in \tab{t:t0}. Since the quark masses are small,
the $\rmO(a)$ correction in \eq{e:betatilde} is not very significant
and it does not matter that we know $b_g$ only to 1-loop. It also does not matter
whether we extrapolate $\hopc$ in $(a/L)^3$ or just use the largest lattice.

Next we need $\lmax/a$ at matching bare couplings $\tilde\beta$. It is found by
interpolating $\tilde\beta=\beta$ such that $\gbar_\mathrm{GF}^2=11.31$ for fixed
$\lmax/a$. Details are referred to \app{s:interpol}. The result is pairs $(\lmax/a,\beta)$. These are subsequently interpolated as $\log(\lmax/a) =P(\beta)$. A linear function $P(\beta)$
does not work well, but second and third order polynomials in $\beta$ do and 
are hardly distinguishable.
We use the second order one and take as uncertainty the typical statistical
error and the difference of the two polynomials added in quadrature.
\Tab{t:t0} contains the results of the first step, at integer values of $\lmax/a$
as well as the numbers interpolated to the CLS bare couplings $\tilde\beta$.
The combination  $({t_0^*})^{-1/2} \lmax$ is listed in the last column of the table.

\begin{figure}[t]
   %\hspace*{-6mm}\includegraphics*[width=1.14\linewidth]{lmaxt0p}
   \hspace*{-0mm}\includegraphics*[width=1.0\linewidth]{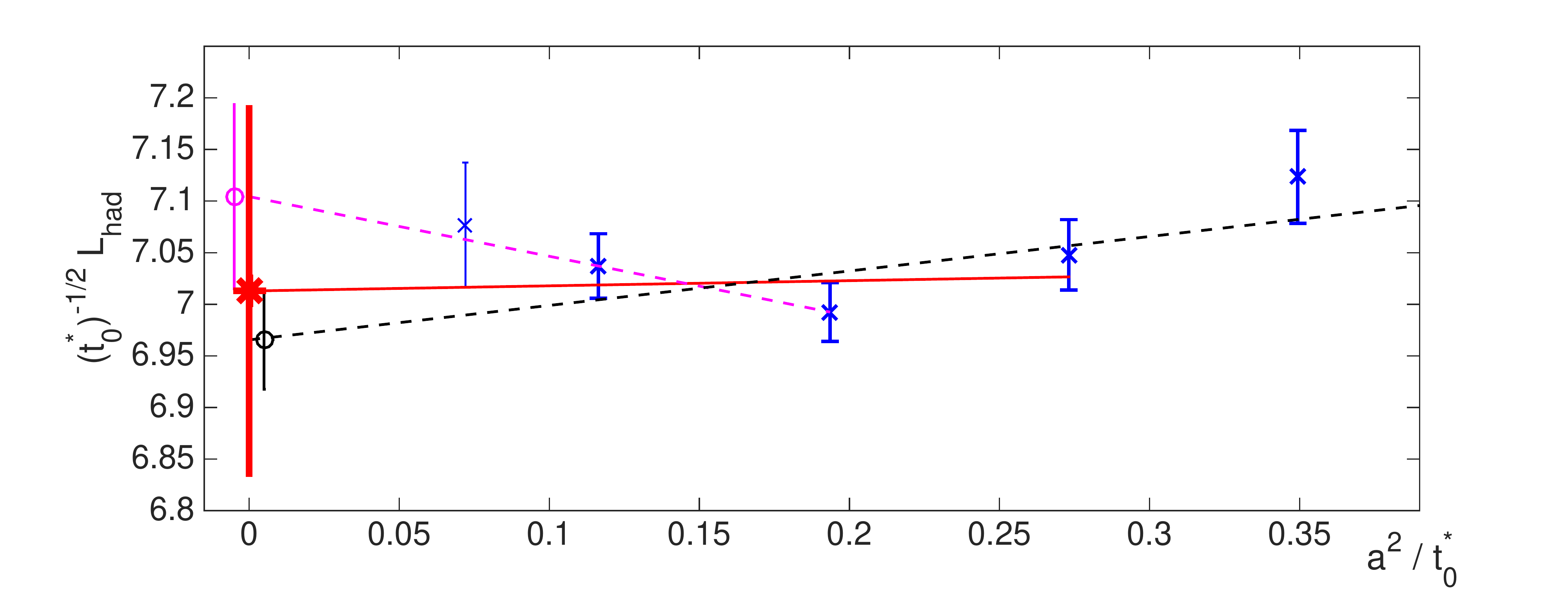}
  \caption{\label{f:lmaxt0} 
  Preliminary continuum extrapolation of  $({t_0^*})^{-1/2}\lmax$.
  The large volume simulation with the smallest lattice spacing is unfinished and
  the correction to shift it to the $\Phi_4=1.11$ point has 
  not yet been included. 
  It is only shown to illustrate where we are heading to. Extrapolations with 
  4,3 and 2 data points are shown together with a range
  for the continuum value covering all of them, see the text. 
  }
\end{figure}

Its continuum extrapolation,
\bes
  ({t_0^*})^{-1/2} \lmax =  \left[({t_0^*})^{-1/2}\lmax \right]_\mathrm{cont} + B\, \frac{a^2}{ t_0^*}\,,
  \label{e:t0lmaxextrap}
\ees
shown in \fig{f:lmaxt0}, is performed with 4,3 and 2 points. 
The preliminary data point at lattice spacing $a=0.04$~fm is not included 
in any of these fits
but rather shown in the graph to illustrate what we will have shortly. 
We take the 3-point extrapolation 
% $\left[({t_0^*})^{-1/2}\lmax \right]_\mathrm{cont}=  7.01(5)$
as our central 
result but enlarge its error of $0.05$ by about a factor four to
\bes
    \left[({t_0^*})^{-1/2}\lmax \right]_\mathrm{cont}=  7.01(18)\,,
\ees
such that it covers the largest 1-sigma excursion 
of all fits, which happens to be the 2-point extrapolation.
It is worth mentioning that the  $a=0.04$~fm 
lattice, as well as all others, is simulated with open boundary conditions in time,
avoiding the freezing of topology~\cite{Luscher:2011kk}. This will allow
very firm conclusions on the continuum limit and a significant reduction of its 
error. 
With the previous numbers we find
\bes
  \lmax= 1.03(3) \,\fm\,, \quad  \Lambda_\msbar^{(3)} = 332(14)\,\MeV\,.
\ees
It is likely that the error of $\lmax$ will shrink significantly once all
preliminary steps are replaced by the final ones.

%%%%%%%%%%%%%%%%%%%%%%%%%%%%%%%%%%%%%%%%%%%%%%%%%%%%%%%%%%%%%%%%%%%%%%%%%%%%%%%
%%%%%%%%%%%%%%%%%%%%%%%%%%%%%%%%%%%%%%%%%%%%%%%%%%%%%%%%%%%%%%%%%%%%%%%%%%%%%%%
%\vspace*{-6mm} 

\section{Connection to the 5-flavor theory and $\alpha_\msbar(m_Z)$}
\label{s:had}
There is little doubt that 3-flavor QCD describes the low energy ($E$)
phenomena including $\Lhad \fpik$ with high precision~\cite{Bruno:2014ufa,Aoki:2016frl}. In other words, the $(E/m_c)^2$ corrections in the effective theory expansion
are small. However,
$\Lambda^{(3)}$ needs to be related to $\Lambda^{(5)}$ because 
physical processes at high energies need 
$\nf\geq5$-flavor QCD and the standard $\alpha_\msbar(m_Z)$ is 
defined in the $\nf=5$ theory.

It has long been known how to connect these theories perturbatively \cite{Weinberg:1980wa,Bernreuther:1981sg} and we now have
4-loop precision \cite{Chetyrkin:2005ia,Schroder:2005hy} in the relation 
\bes
 \gbar^{(\nf-1)} (m_*) = \gbar^{(\nf)} (m_*) (1 + \rmO([\gbar^{(\nf)}(m_*)]^4)\,,
 \label{e:gnfm1}
\ees
where $m_*=\mbar_\msbar(m_*)$ is the mass of the decoupled quark
in the $\nf$-flavor theory and in the $\msbar$-scheme. 
Together with 
\begin{eqnarray}
    \Lambda^{}_s &=& \varphi^{}_s(\gbar^{}_s(\mu))\, \times\, \mu\,,
    \label{e:Lam} 
    \\
    \varphi_s(\gbar_s) &=& ( b_0 \gbar_s^2 )^{-b_1/(2b_0^2)} 
        \rme^{-1/(2b_0 \gbar_s^2)} \label{e:phig} \,
        \times\, \exp\bigg\{-\int\limits_0^{\gbar_s} \rmd x\ 
        \Big[\frac{1}{\beta_s(x)} 
             +\frac{1}{b_0x^3} - \frac{b_1}{b_0^2x} \Big] \bigg\} \,. 
        \nonumber                                
\end{eqnarray}
$\beta\to\beta^\mathrm{pert}_\msbar$, we can compute the ratio of the $\Lambda$-parameters at given values of $m_\star$. The b- and c-quark masses,
$m_\star=4.18\,\GeV$ and $m_\star=1.275\,\GeV$ are 
taken from the PDG~\cite{Agashe:2014kda}. 
With the available perturbative precision~\cite{MS:4loop1,Czakon:2004bu,Baikov:2016tgj,Luthe:2016ima,Chetyrkin:2005ia,Schroder:2005hy}, we find
% \bes
%     \Lambda_\msbar^{(4)}=289(14)\,\MeV\,,\; \Lambda_\msbar^{(5)}=207(11)\,\MeV
%\ees
%and
%\bes
%   \alpha_\msbar(m_Z)=0.1179(10)(2)\,. 
%   \label{e:alphamz}
%\ees
\bes
     \Lambda_\msbar^{(4)}&=&289(14)\,\MeV\,,\;\;
     \Lambda_\msbar^{(5)}= 207(11)\,\MeV\,,\; 
      \\[-0.5ex]
    \alpha_\msbar(m_Z)&=& 0.1179(10)(2)\,. 
   \label{e:alphamz}
\ees
The first error in $\alpha$ is just propagated from the one in $\Lambda_\msbar$, 
which in turn is obtained by standard error propagation of all previously 
discussed numbers which were put together. The second error represents 
our estimate of the uncertainty from using PT in the connection
$\Lambda_\msbar^{(3)}\to\Lambda_\msbar^{(5)}$. We arrive at it as follows.
The $2,3,4$-loop terms in \eq{e:gnfm1} combined
with the  $3,4,5$-loop running lead to contributions
$109,\,15,\,7$
(in units of $10^{-5}$) to  $\alpha_\msbar(m_Z)$. 
We take the sum of the last two contributions 
as our error in \eq{e:alphamz}. {\em Within PT}, this represents a very conservative error estimate: the known terms of the series behave
similar to a convergent series but we treat it like an asymptotic one.

However, we have to stress that we are here using perturbation
theory at the scale of the charm quark mass. In principle it is
possible that PT is entirely misleading
when we apply it at such low scales, decoupling the charm quark. 
One may note that almost all lattice determinations as well as
a number of continuum ones have this same error, but this does not help much. 
As long as we do not have a computation of all the above steps with $\nf=4$,
we have to live with our estimate in \eq{e:alphamz} and 
with this -- in our opinion unlikely \cite{Bruno:2014ufa} -- possibility. 
It would mean that the second error estimate is far off due to a breakdown of PT for 
 $\Lambda^{(3)}/\Lambda^{(4)}$.

\input{acknow}

\appendix
\section{Interpolations to $\gbar_\mathrm{GF}^2=11.31$\label{s:interpol}}
The point of reference where we match between the hadronic world 
and the finite volume GF coupling is 
\bes
   \gbarGF^2(\lmax) = 11.31 \,,\quad m(\lmax)=0\,.
   \label{e:lcp}
\ees
We discuss the present, preliminary, interpolations of the available 
coupling data to this point separately in this appendix 
because it is rather technical and the technical difficulties
mostly are due to the presently incomplete set of simulation data.
This will change soon.

The difficulty is that one needs to have the quark mass set to zero in a 
precisely defined way, with a fixed condition as one varies the lattice spacing.
We need a unique line of constant physics. Then cutoff effects are smooth 
functions of $a$ with the asymptotic form of \eq{e:t0lmaxextrap}. 
\Eq{e:lcp} is the natural condition, where $m$ is the improved PCAC mass
in a $L^4$ lattice with the same Dirichlet boundary conditions
as in the definition of $\gbarGF$. More details are found in \cite{DallaBrida:2016kgh,Nf3tuning}.

Unfortunately, the presently available data for $\gbarGF^2 = F(L/a,\beta)$ 
do not homogeneously satisfy $m(L)=0$.
For $L/a=12$ they do.  But on the larger lattices, $L/a=16,24,32$,
we have $m(L/2)=0$,
because they originate from the computation of step scaling functions 
\cite{DallaBrida:2016kgh}. There is a $\rmO((a/L)^2)$ cutoff effect
between the two definitions. We checked for its size:
we interpolated the data with $L/a=16$,  $m(L/2)=0$ in $\beta$ to $\gbarGF^2 = 11.31$ finding $\beta=3.5607$. 
At this $\beta$, a computation adjusted such that 
$m(L)=0$ yields $\gbarGF^2 = 11.03(6) = 11.31-\Delta g^2$. Presently, we take this effect into account
by treating it as small and in lowest order: we modify \eq{e:lcp} 
for $L/a \geq 16$ to $\gbarGF^2(\lmax) = 11.31 + \Delta g^2 \frac{16^2}{(L/a)^2}$ at $m(\lmax/2)=0$.
The $L/a,\beta$ points satisfying this condition are found by a 
quadratic interpolation of the form $F(L/a,\beta) = [k_0 + k_1\beta +k_2\beta^2]^{-1}$ for fixed $L/a$
implemented by a fit to about 5 data points in the vicinity. 
The resulting pairs $(\lmax/a,\beta) = (\lmax/a,\tilde\beta)$ are listed in 
\tab{t:t0}.

\bibliographystyle{JHEP}
\bibliography{gbarlett,qcdpt,GFpaper,scalesett,main}

%\begin{thebibliography}{99}
%\bibitem{...}
%....

%\end{thebibliography}

\end{document}

%% file: talk_sint.tex
\section{Introduction}

The strong coupling in QCD is a fundamental parameter of the Standard Model and 
its precise knowledge is both of principal importance and of practical relevance to LHC physics. 
We here report on the ALPHA collaboration's results for the $\Lambda$-parameter in 3-flavour
QCD and $\alpha_s(m_Z)$. The project was designed to match hadronic quantities computed
on CLS gauge configurations for 3-flavour QCD at low energies~\cite{Bruno:2014jqa,Bruno:2016plf}.
The strategy for the $\Lambda$-parameter was explained in~\cite{Brida:2014joa,Brida:2015gqj} 
and relies on the methods and tools developed over many 
years~(see \cite{Luscher:1998pe,Sommer:2006sj} and references therein).
As a result we are able to defer the use of perturbation theory in 3-flavour QCD 
to high energies of O(100 GeV). On the other hand, the determination of $\alpha_s(m_Z)$ requires
the matching to the 4- and 5-flavour theories across the charm and bottom quark thresholds 
which still relies on perturbation theory~\cite{Bernreuther:1981sg,Chetyrkin:2005ia}.
At all stages the continuum limit is taken and rather well controlled. 
Our strategy combines the perturbative knowledge of the standard SF coupling at high energies
with the advantageous properties of finite volume gradient flow
couplings at low energies. A non-perturbative matching between these 2 coupling schemes is performed at 
an intermediate scale $1/L_0\approx 4\,{\rm GeV}$. At the largest box size reached, $\lmax$,
the matching to the gradient flow time scale $t_0^\ast$ at the SU(3) symmetric point is performed. 
Together with the recent results of ref.~\cite{Bruno:2016plf} (which rely on a new high
precision determination of the axial current normalization constant~\cite{brida2016}, based on
the method in~\cite{Brida:2016rmy}), this allows us to accurately relate to a linear combination of pion and kaon
decay constants and thus express all results in physical units.

In this write-up we go through the different steps of this strategy starting from the high
energy end (section 2). The connection between the intermediate scale $L_0$
and $\lmax$ is discussed in section 3, including the matching at scale $L_0$. 
Relating $\lmax$ to a hadronic scale is the subject of section~4. This allows
to quote the 3-flavour $\Lambda$-parameter in physical units. The perturbative
connection to 5-flavour QCD is carried out in section~5, followed by our conclusions. 
Finally, a technical point pertaining
to the interpolation of the low energy data is relegated to an appendix.
Note that the first 2 steps have been published in \cite{Brida:2016flw} and \cite{DallaBrida:2016kgh}, respectively. 
Further details on the first step will be given elsewhere~\cite{SFcoupinpreparation}.
The matching of $\lmax$ to a hadronic scale is currently being finalized.
For a recent account aimed at a non-lattice audience we refer to~\cite{Bruno:2016gvs}.

\section{The high energy regime}
\label{s:sint}

\subsection{A family of couplings in the SF scheme}
Using the Schr\"odinger functional in QCD~\cite{Luscher:1992an,Sint:1993un}, the spatial vector components of the gauge field 
at the time boundaries $x_0=0,T$ are taken to be
spatially constant and Abelian~\cite{Luscher:1993gh},
\begin{subequations}
  \begin{eqnarray}
A_k(x)\big|_{x_0=0} = C_k^{} &=& \dfrac{i}{L}{\rm diag}\left(\eta-\frac{\pi}{3},\eta\left(\nu-\frac12\right),
                                        -\eta\left(\nu+\frac12\right)+\frac{\pi}{3}\right),\\
A_k(x)\big|_{x_0=T} = C_k'   &=& \dfrac{i}{L}{\rm diag}\left(-\pi-\eta,\eta\left(\nu+\frac12\right)+\frac{\pi}{3},
                                        -\eta\left(\nu-\frac12\right)+\frac{2\pi}{3}\right),
\end{eqnarray}
\label{eq:ck}
\end{subequations}
for $k=1,2,3$. The parameters $\eta$ and $\nu$ correspond 
with the existence of 2 abelian generators in SU(3).
The absolute minimum of the action with these boundary values is attained for~\cite{Luscher:1992an}
\begin{equation}
    B_k(x) = C_k + \dfrac{x_0}{T}\left(C_k'-C_k\right),\qquad B_0=0\,,
\end{equation}
the induced background field, which is unique up to gauge equivalence. 
The effective action of the Schr\"odinger functional
$\Gamma[B]$ is then unambiguously defined and its perturbative expansion 
straightforward in principle,
\begin{equation}
   \Gamma[B] = \frac{1}{g_0^2}\Gamma_0[B] + \Gamma_1[B] + {\rm O}(g_0^2),
\end{equation}
with the first term given by the classical action of the background field,
\begin{equation}
   \Gamma_0[B]= g_0^2 S[B] =2(\pi+3\eta)^2\,.
\end{equation}
Setting all quark masses to zero and $T=L$, the only remaining scale is set by $L$.
The SF coupling at this scale is then defined as a derivative 
with respect to a background field parameter,
\begin{equation}
    \left.\dfrac{\partial_\eta \Gamma[B]}{\partial_\eta \Gamma_0[B]}\right\vert_{\eta=0}
   = \dfrac{\left.\left\langle \partial_\eta S \right\rangle\right\vert_{\eta=0}}{12\pi}
   = \dfrac{1}{\bar{g}^2(L)} - \nu\times \bar{v}(L)\,.
   \label{eq:gnudef}
\end{equation}
The derivative produces an expectation value which can be measured in numerical simulations.
In fact, there are 2 observables, the inverse coupling $1/\bar{g}^2(L)$ and $\bar{v}(L)$ both of which are
measured at $\nu=0$. A new feature of our project is the re-interpretation of the parameter
$\nu$ as index of a family of SF couplings,
\begin{equation}
   \dfrac{1}{\bar{g}_\nu^2(L)} = \dfrac{1}{\bar{g}^2(L)} - \nu\times \bar{v}(L) \,.
\end{equation}
This has first been envisaged as a method to reduce large cutoff effects in 
strongly coupled models of electroweak symmetry breaking in~\cite{Sint:2012ae}.

In perturbation theory the $\beta$-function of the SF coupling family is known to 3-loops from the 2-loop matching to 
the $\overline{\rm MS}$-coupling in \cite{Bode:1999sm,Luscher:1995nr,Christou:1998wk,Christou:1998ws} combined
with the knowledge of the $\beta$-function in the 
$\overline{\rm MS}$-scheme~(cf.~\cite{MS:4loop1,Czakon:2004bu} and references therein),
\begin{equation}
   \beta(\bar{g}_\nu^{}) = - L \frac{\partial \bar{g}_\nu^{}(L)}{\partial L}  
   = -b_0  \bar{g}_\nu^{3} -b_1 \bar{g}_\nu^{5} -b_{2,\nu} \bar{g}_\nu^{7} + {\rm O}(\bar{g}_\nu^{9})\,.
\end{equation}
In 3-flavour QCD the universal terms are $b_0= 9/(4\pi)^2 $  and $b_1=1/(4\pi^4)$. 
The 3-loop coefficient is then found to be
\begin{equation}
   b_{2,\nu} = \left(-0.06(3) -\nu\times 1.26\right)/(4\pi)^3\,.
\end{equation}
The $\Lambda$-parameter in the ${\rm SF}_\nu$ scheme is defined by
\begin{equation}
   L\Lambda_{{\rm SF}_\nu} = \varphi_\nu\left(\bar{g}_\nu^{}(L)\right) 
   = \big[{b_0\bar{g}_\nu^2(L)}\big]^{-\frac{b_1}{2b_0^2}}\,{\rm e}^{-\frac{1}{{2b_0\bar{g}_\nu^2(L)}}}
      \exp\bigg\{-\int_0^{\bar{g}_\nu^{}(L)}\! dg
    \bigg[\dfrac{1}{\beta(g)}+\dfrac{1}{b_0g^3} -\dfrac{b_1}{b_0^2g}\bigg]\bigg\}\,.
    \label{eq:L0timesLambda}
\end{equation}
When evaluating the $\Lambda$-parameter one would like to know from which scale $L$ one can
trust perturbation theory to evaluate the integral in the exponent 
and how one can quantify the associated systematic error.
We proceed as follows: first we fix a reference scale, $L_0$, through
\begin{equation}
  \bar{g}^2(L_0) = 2.012,
\end{equation}
and use the step-scaling function (cf.~next subsection) to step up the energy scale
non-perturbatively by factors of 2 from $L_0$ to $L_n=L_0/2^n$, for $n=0,1,2,\ldots$.
We then use the $\nu=0$ $\Lambda_{\rm SF}$-parameter as reference point and consider
\begin{equation}
   L_0\Lambda_{\rm SF}  = \underbrace{(\Lambda_{\rm SF} /\Lambda_{{\rm SF}_\nu})}_{\exp(-\nu\times1.25516)} 
   \times \underbrace{(L_0/L_n)}_{2^n} \times \varphi_\nu(\bar{g}_\nu^{}(L_n))
\end{equation}
with $\varphi_\nu(\bar{g}_\nu^{}(L_n))$ evaluated in perturbation theory, by inserting the value of 
the coupling $\bar{g}_\nu^2(L_n)$ obtained non-perturbatively from the step-scaling procedure
and the 3-loop truncated $\beta$-function.
Up to perturbative errors of order $\alpha^2(1/L_n)=\bar{g}_\nu^4(L_n)/(4\pi)^2$, the result for $L_0\Lambda$ must
be independent of the number of steps $n$ and the value of the parameter $\nu$. This gives
us an excellent control over the remaining systematic error stemming from perturbation theory.
Before discussing the result we briefly present some key features of our numerical simulations, 
the measurements and the data analysis.

\subsection{Numerical simulation data and analysis}
\begin{figure}[t]
\centering
\includegraphics[width=9cm,clip]{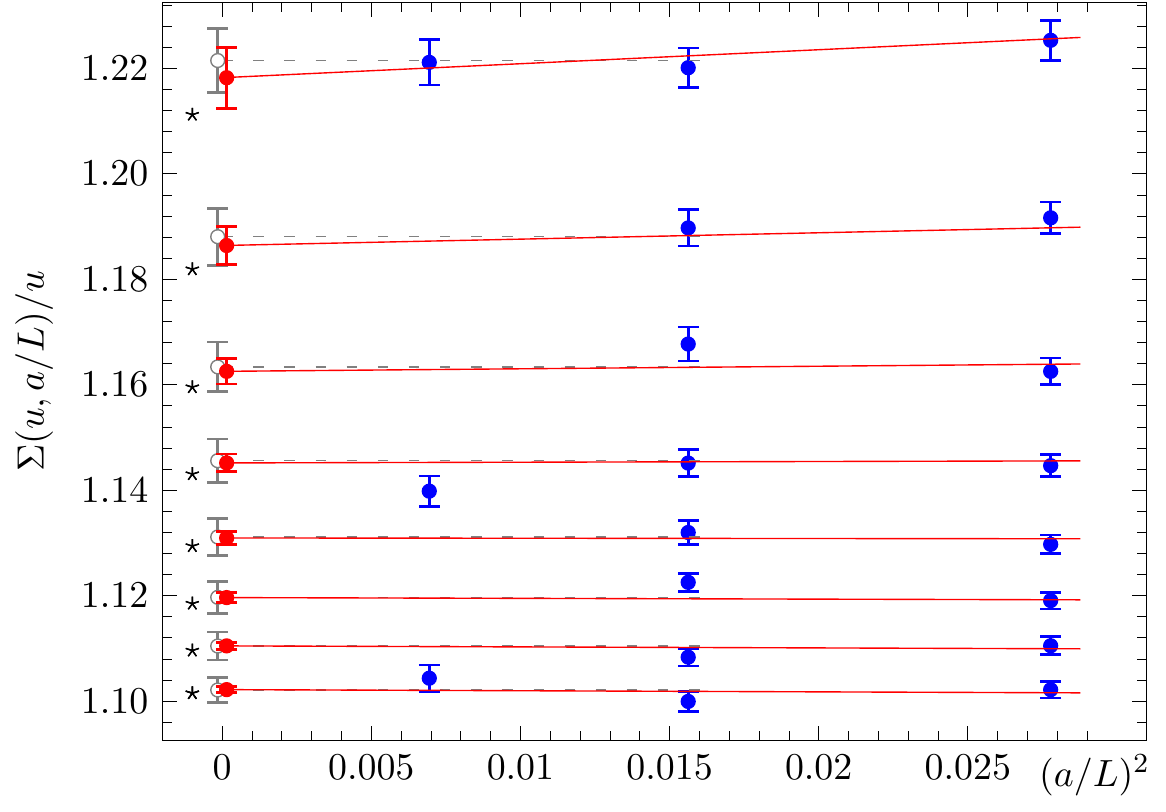}
\caption{Continuum extrapolation of the step scaling function. 
The leftmost points are the continuum values, whereas
the stars are obtained from perturbative scale evolution using the 3-loop $\beta$-function.}
\label{fig:CLssf}
\end{figure}
For the simulation at high energies we choose the Wilson plaquette action in order to use
2-loop perturbative information at finite $L/a$ which is only available for
this regularization~\cite{Bode:1999sm}. 
This allows for more control of boundary O($a$) effects and also for perturbative
improvement of the non-perturbative data of the step-scaling function.
We use the result for $\csw$ from \cite{Yamada:2004ja}. A very careful tuning of the bare
mass parameters to their critical values in \cite{Nf3tuning}
reduces associated systematics to negligible levels.
All our simulations have been carried out with a modification of
the openQCD code~\cite{Luscher:2012av}.
Compared to earlier studies of the SF coupling in 3-flavour QCD in \cite{Aoki:2009tf} (there with the Iwasaki gauge action),
we have significantly reduced statistical errors.
We have produced data for the step-scaling functions 
\begin{equation}
  \Sigma_\nu(u,a/L) = \bar{g}_\nu^2(2L)\vert_{\bar{g}_\nu^2(L)=u,m(L)=0}\,,
\end{equation}  
for lattice resolutions $L/a=6,8,12$ and the corresponding doubled 
lattice sizes\footnote{For $L/a=12$ we have limited the data production on the 24-lattices to 3 parameter choices.}.
For $\nu=0$ our data corresponds to $u$-values in the interval $[1.1,2.0]$.
Cutoff effect are O($a^2$) in the bulk, and O($a$) from the time boundaries. The latter are
governed by 2 counterterms of dimension 4 with coefficients $c_{\rm t}$ and $\tilde{c}_{\rm t}$, known to 
2-loop~\cite{Bode:1999sm} and 1-loop order~\cite{Luscher:1996vw} respectively. 
Using these perturbative results for the coefficients
we expect that O($a$) effects are strongly reduced. As a safeguard against any
O($a$) contamination of our continuum extrapolations we include a systematic
error as follows. We determine the $c_{\rm t}$ and $\tilde{c}_{\rm t}$ derivatives of the coupling
numerically by a variation in a few points and, together with the corresponding
perturbative information arrive at a model for the sensitivity of our data to
such variations. We then use the last known perturbative coefficients for $c_{\rm t}$ and $\tilde{c}_{\rm t}$ 
as an uncertainty and add the corresponding systematic error to the data. Note that this is likely
an overestimate of the true error~\cite{SFcoupinpreparation}. 
Fig.~\ref{fig:CLssf} shows our data points for the step-scaling function at $\nu=0$.
We then perform global fits of the data, for example
\begin{equation}
\Sigma^{(2)}(u,a/L) = u + s_0 u^2 + s_1 u^3 
+  { c_1} u^4 + { c_2} u^5 
+  {\rho_1} u^4 \dfrac{a^2}{L^2} + {\rho_2} u^5 \dfrac{a^2}{L^2}\,,
\end{equation}
where $\Sigma^{(2)}$ denotes the 2-loop improved non-perturbative data,
such that cutoff effects appear, by construction, first at O($u^4$)~\cite{deDivitiis:1994yz}.
The coefficients $s_0$, $s_1$ are fixed to their perturbative values,
\begin{equation}
  s_0 = 2b_0\ln 2,\quad  s_1 = s_0^2 + 2b_1\ln 2\,,
\end{equation} 
and we thus obtain the non-perturbative continuum step-scaling function $\sigma(u)$ 
in terms of the fit coefficients $c_1$ and $c_2$ and their correlation, as required for
the error propagation.

\subsection{Results for $L_0\Lambda$}
\begin{figure}[h]
\centering
\includegraphics[width=9cm,clip]{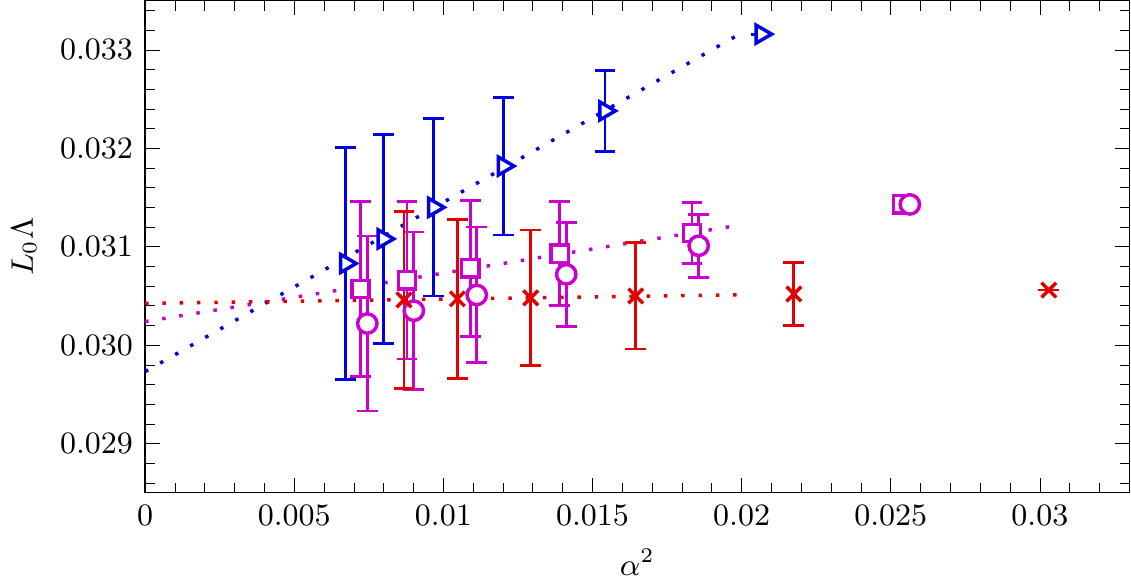}
\caption{The extraction of the $\Lambda$-parameter using perturbation theory at different values of $\alpha$, 
plotted vs. $\alpha^2(1/L_n)$. The data points are, from top to bottom, 
for $\nu=-0.5,0,0.3$ and, from right to left, for $n=0,1,\ldots,5$ steps by a factor 2 in scale.}
\label{fig:LambdaL0}       % Give a unique label
\end{figure}
Given the step-scaling functions in the continuum limit our data allows us to take up to $n=5$ steps
from $L_0$, thus covering a factor of $2^5=32$ in scale. For $\nu\ne 0$ one also needs
to know the value of $\bar{v}$ at $L_0$~\cite{Brida:2016flw},
\begin{equation}
 \bar{v}(L_0) = 0.1199(10)\,.
 \label{eq:vbarL0}
\end{equation}
We have considered  O(10) $\nu$-values around $\nu=0$. We present data for $\nu=0.3$ and $\nu=-0.5$ 
besides the reference choice $\nu=0$. 
Fig.~\ref{fig:LambdaL0} shows the corresponding $3\times 6$ data points for $L_0\Lambda$ from eq.~(\ref{eq:L0timesLambda}),
each corresponding to a determination of $L_0\Lambda$ using non-perturbative data between 
$L_0$ and $L_n$ and perturbation theory for $L<L_n$. As can be seen in the plot the data
points nicely come together at small $L$ (high energy scales) where $\alpha(1/L_n)\approx 0.1$.
We are therefore confident to quote our result with an error of $3\%$,
\begin{eqnarray} 
 L_0\Lambda_{\rm SF}  =0.0303(8)  \quad \Leftrightarrow \quad
 L_0 \Lambda_\msbar^{(3)} = 0.0791(21)\,,  
 \label{e:llresult2}
\end{eqnarray} 
which is the main result of our study of the large energy region. 
In a recent letter~\cite{Brida:2016flw} we have also presented a
by-product of this study, namely the observation that renormalized perturbation theory
in continuum QCD might be susceptible to larger systematic errors than often assumed.
This is apparent in fig.~\ref{fig:LambdaL0} and in the  continuum result for $\bar{v}(L)$ where perturbation
theory at $\alpha=0.19$ does not look trustworthy. This is particularly worrisome given the
very advantageous properties of the ${\rm SF}_\nu$-schemes in perturbation theory~\cite{Brida:2016flw,SFcoupinpreparation}.

%% file: talk_ramos.tex
\section{Connecting with hadronic scales}

In the previous section we have detailed our very accurate matching
with perturbation theory.  The result is given in eq.~(\ref{e:llresult2})
%\begin{equation}
%  \Lswi \Lambda^{(3)}_\mathrm{\overline{MS}} =0.0791(21)\,,
%\end{equation}
and relates the $\Lambda$-parameter with $L_0$, 
defined implicitly through $\gbar^2_{\rm SF}(L_0) = 2.012$.
The corresponding energy  scale, $1/L_0\approx 4{\rm GeV}$, 
is still very large if one aims at using lattice methods to
study continuum properties  of QCD while having finite volume effects under control. 
Therefore we will continue using our finite size scaling technique in order to relate
the scale $L_0$ with a scale $\lmax$ characteristic of hadronic
physics. 

In principle one could simply continue with the program explained in
the previous section until reaching the energy scale $1/L_{\rm
  had}$. Unfortunately the statistical 
precision of the SF coupling deteriorates very fast when reaching
large volumes (see for example the discussion
in~\cite{Ramos:2015dla} and references therein), making it very
difficult to maintain the precision. In order to overcome these
problems we will continue to work with a
finite volume renormalization scheme, but we will use the 
gradient flow coupling in a finite volume with SF boundary
conditions~\cite{Fritzsch:2013je} for our running
coupling. Renormalized couplings based on the gradient flow have the
nice property that their variance is roughly independent of the lattice
spacing. 

The main result of this section~\cite{DallaBrida:2016kgh} is the
precise determination of the ratio of two scales
\begin{equation}
  \Lhad/L_0 =  21.86(42) \quad \text{for} \quad \gbar^2_{\rm
    GF}(\Lhad)=11.31\,. 
  \label{e:lhadl0}
\end{equation}
Note, however, that in the usual step scaling procedure (see previous
section) the aim is to 
determine the value of the renormalized coupling at two scales $L_1$
and $L_2$ that are separated by an integer power of the scale factor
(i.e. $L_1 = 2^nL_2$ with $n\in \mathbb Z$ when the scale factor is
$s=2$). Here we need to solve a 
slightly different problem: we know the value of the coupling at two
scales ($L_0$ and $\lmax$), and are interested in computing the
ratio of these scales. With our conventions\footnote{We recall that we
  always work in a mass-independent
renormalization scheme, and therefore the $\beta$-function only
depends on $g$.}, the $\beta$-function is
defined by 
\begin{equation}
  -L\frac{\partial \gbar_{\rm GF}(L)}{\partial L} = \beta(\gbar_{\rm GF})\,,
\end{equation}
and ratios of scales such as~\eqref{e:lhadl0} can be easily computed,
\emph{once the $\beta$-function is known}, thanks to the relation
\begin{equation}
  \frac{L_1}{L_2} = \exp\left\{
    -\int_{\gbar_{\rm GF}(L_2)}^{\gbar_{\rm GF}(L_1)} \frac{dx}{\beta(x)}
  \right\}\,.
\end{equation}
In order to determine the $\beta$-function, we will first determine
the usual continuum step scaling function,
\begin{equation}
  \sigma_{\rm GF}(u) = \gbar^2_{\rm GF}(2L)\Big|_{g^2_{\rm GF}(L)=u}\,,
\end{equation}
and use it to constrain the functional form of $\beta(g)$ by
using the exact relation
\begin{equation}
  \log 2 =  -\int_{\sqrt{u}}^{\sqrt{\sigma_{\rm GF}(u)}} \frac{dx}{\beta(x)}\,.
\end{equation}

\subsection{Coupling definition and choices of discretization}

The gradient flow coupling with SF boundaries conditions has been
studied in~\cite{Fritzsch:2013je}, and the interested reader should
consult the cited reference. We impose SF boundary conditions with
zero background field (i.e. $C_k = C_{k'} = 0$ in eq.~\eqref{eq:ck}). The
gradient flow~\cite{Narayanan:2006rf,Luscher:2010iy} determines how
the gauge field $B_\mu(t,x)$ evolves with the flow time $t$ (not to be
confused with the Euclidean time $x_0$) via the (non-linear) diffusion-like
equation 
\begin{equation}
  \label{eq:flow}
  \frac{\partial B_\mu(t,x)}{\partial t} = 
  D_\nu G_{\nu\mu}(t,x);\quad A_\mu(x) = B_\mu(t,x)\Big|_{t=0}; \quad 
        G_{\mu\nu} = \partial_\mu B_\nu - \partial_\nu B_\mu + [B_\mu,B_\nu]\,.
\end{equation}
The important point is that gauge invariant observables made out of
the flow field $B_\mu(t,x)$ are automatically
renormalized~\cite{Luscher:2011bx} for $t>0$. In particular one can use the
action density ${\rm Tr}(G_{\mu\nu}G_{\mu\nu})$ to define a
renormalized coupling at a scale $\mu \propto 1/\sqrt{8t}$. There are many
subtleties that lead us to use the following coupling definition
\begin{equation}
  \gbar^2_{\rm GF}(L) = \mathcal N^{-1}\frac{t^2}{4}\sum_{i,j=1}^3
  \frac{\langle{\rm Tr}(G_{ij}(t,x)G_{ij}(t,x)) \hat\delta(Q)\rangle}
  {\langle\, \hat\delta(Q)\,\rangle}\Big|_{\sqrt{8t}=cL;\, x_0=T/2;\,T=L}\,.
\end{equation}
Note that the renormalization scale runs with the volume thanks to the
relation $\sqrt{8t}=cL$ (We choose $c=0.3$ in this work). Several points
require some explanation 
\begin{enumerate}
\item The SF breaks the invariance under translations in Euclidean
  time. Moreover full ${\rm O}(a)$ improvement with the SF setup
  requires to determine non-perturbatively two boundary improvement
  coefficients (i.e. $c_{\rm t}, \tilde{c}_{\rm t}$), which 
  we only know to 1-loop order
  in perturbation theory (see below). To minimize these boundary effects,
  we choose to define the coupling using the action density at
  $x_0=T/2$, and based only on the magnetic components ($i,j=1,2,3$) 
  of the field strength tensor. With
  these choices boundary ${\rm O}(a)$ effects can be estimated from our
  data set, and are found to be small.

\item Simulations with SF boundary conditions suffer from the common
  problem of topology freezing~\cite{DelDebbio:2004xh}. In order to
  overcome this problem, we define the coupling only in the sector with zero
  topological charge (see~\cite{Fritzsch:2013yxa} for more details).  
  
\item For the precise definitions of the normalization $\mathcal N$ and
  the delta-function $\hat\delta(Q)$ projecting to the zero charge sector
  we refer to \cite{DallaBrida:2016kgh}.
\end{enumerate}
Since our final aim is to determine our scales in physical units from
the large volume simulations of the CLS initiative~\cite{Bruno:2014jqa}, 
we use the same bare lattice action, except for the Euclidean time  
boundary conditions. In particular, we
simulate three massless flavours of non-perturbatively ${\rm O}(a)$
improved Wilson fermions and a Symanzik tree-level ${\rm O}(a)$ improved
gauge action. There is some freedom when defining this action near
the Euclidean time boundaries, and we use \texttt{option B} of
reference~\cite{Aoki:1998qd}. With this choice we know the 1-loop value of
the boundary improvement coefficients $c_{\rm t},
\tilde{c}_{\rm t}$~\cite{Takeda:2003he,pol}.

There are many possibilities when translating the continuum flow
equation~\eqref{eq:flow} to the lattice. Different definitions lead to
different cutoff effects, but these can be
large. Following ref.~\cite{Ramos:2015baa} we choose the
\emph{Zeuthen flow}. This particular discretization guarantees that
${\rm O}(a^2)$ cutoff effects are not generated when integrating
the flow equation. In~\cite{DallaBrida:2016kgh} we performed a
detailed study of the cutoff effects of flow quantities and concluded
that, at least for our data set, the scaling properties of the
\emph{Zeuthen flow} allow us to perform a more accurate continuum
extrapolation.

\subsection{Determination of the step scaling function}

The bare mass $m_0$ is tuned to its critical value $m_{\rm cr}$
with excellent precision~\cite{Nf3tuning},
for all values of the bare coupling
$\beta=6/g_0^2$ used here. Any deviation
from the critical line is well below our 
statistical uncertainties. Given the function $m_{\rm
  cr}(\beta,L/a)$ (see~\cite{Nf3tuning}), our simulations depend
essentially only on one bare parameter $\beta$. 

In order to determine the step scaling function we perform 9 precise
simulations at $\beta \in \{3.9,4.0,4.1,4.3,4.5,4.8,5.1,5.4,5.7\}$ on a
$L/a=16$ lattice. These
simulations provide our 9 target couplings $v_i$ 
\begin{equation}
  v_i \in \{2.1257, 3.3900, 2.7359, 3.2029, 3.8643, 4.4901, 5.3013,
    5.8673, 6.5489\},
\end{equation}
for which we would like to obtain the continuum step scaling function $\sigma(v_i)$. In
order to do this, we tune the bare coupling $\beta$ on some $L/a=8,12$
lattices to match these values of the renormalized
coupling. Once this tuning is satisfactory, we compute 
$\gbar^2_{\rm GF}$ on the doubled lattices ($L/a=16,24,32$), with all
bare parameters kept fixed, and thus obtain the
lattice step scaling function $\Sigma_{\rm GF}(v_i, L/a)$.

We find that our non perturbative data follows very closely the
functional form
\begin{equation}
  \label{eq:prop}
  \frac{1}{\Sigma_{\rm GF}(u,L/a)} - \frac{1}{u} = \text{constant}\,,
\end{equation}
which allows us to propagate the uncertainties in $u$ into
$\Sigma_{\rm GF}$ by using $\partial \Sigma_{\rm GF}/\partial u =
\Sigma_{\rm GF}^2/u^2$. A 
(conservative) estimate of the boundary effects due to $c_{\rm t}$ is also
propagated to $\Sigma_{\rm GF}$ in a similar fashion
(see~\cite{DallaBrida:2016kgh} for the details).  

\begin{figure}[h]
  \centering
  \begin{subfigure}[t]{0.48\textwidth}
    \centering
    \includegraphics[width=\textwidth]{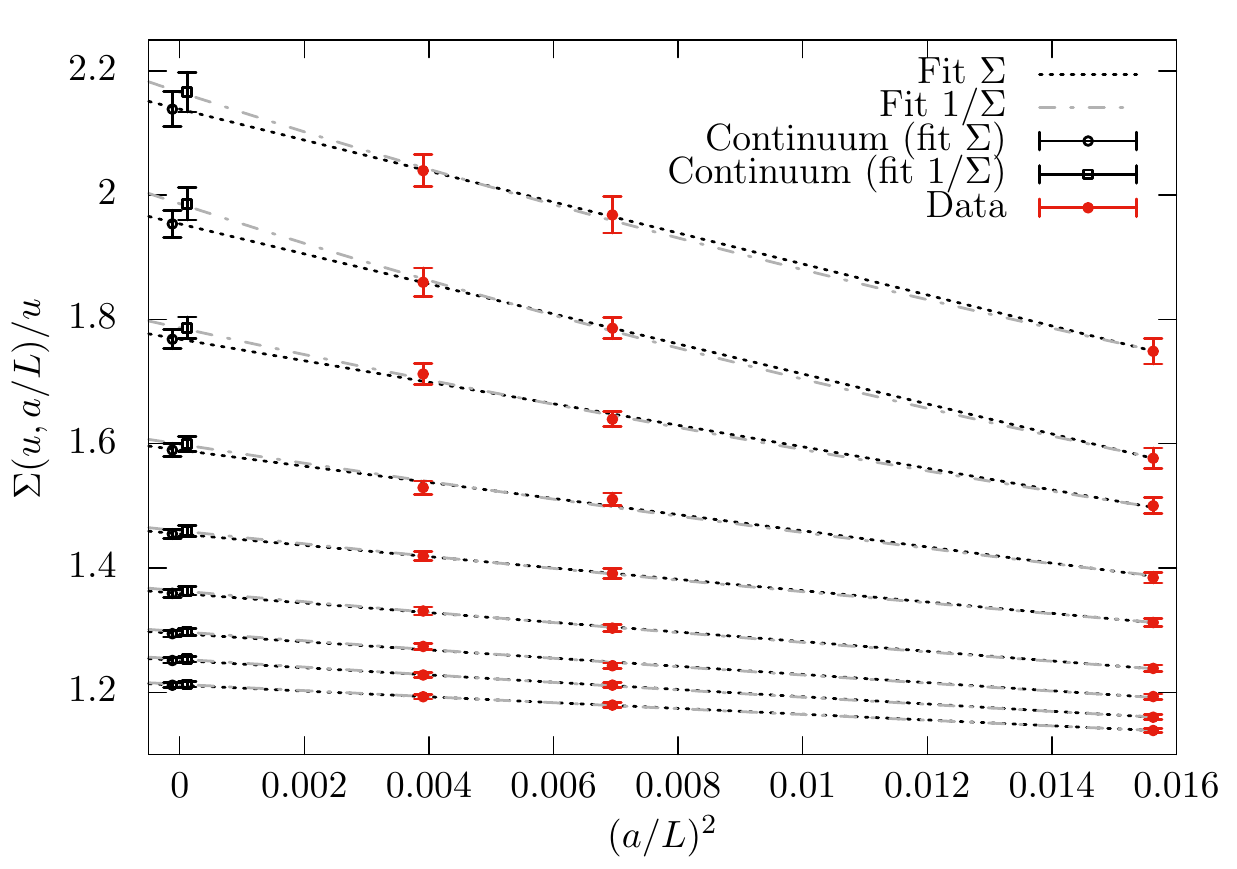}
    \caption{Continuum extrapolation of the step scaling function at the
      9 target values of the coupling using as weights for the fit the
    uncertainty in $\Sigma$. Note that different continuum
    extrapolations are systematically different.}
    \label{fig:sigma}
  \end{subfigure}
  \centering
  \begin{subfigure}[t]{0.48\textwidth}
    \centering
    \includegraphics[width=\textwidth]{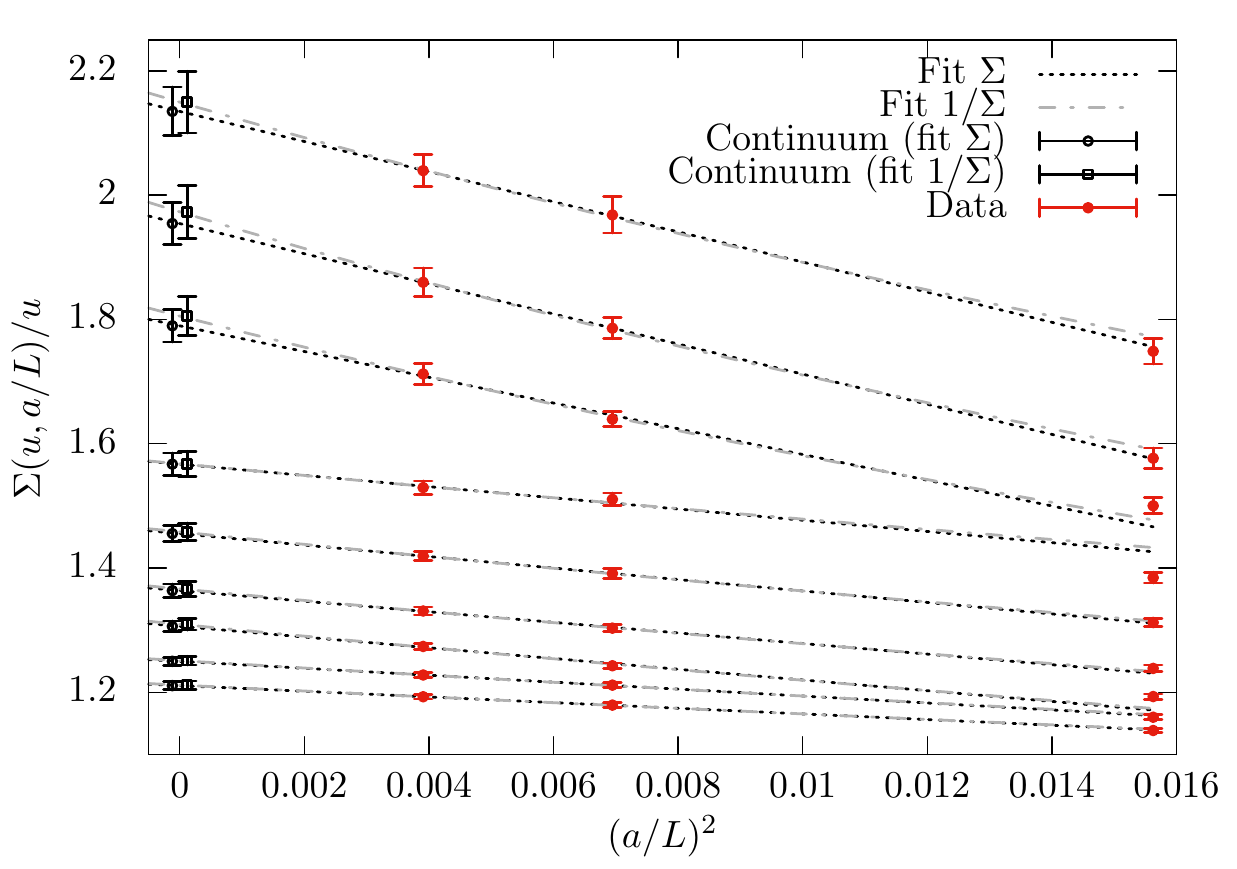}
    \caption{Continuum extrapolation of the step scaling function at the
      9 target values of the coupling using as weights for the fit the
    uncertainty in $\Sigma$ and the systematic estimate of the 
    ${\rm O}(a^4)$ effects.}
    \label{fig:sigma_a4}
  \end{subfigure}
  \caption{Continuum limit of the step scaling function with and
    without including the systematic effect~\eqref{eq:sys}.}
    \label{fig:sigmas}
\end{figure}

All in all we get lattice estimates of the step scaling function
$\Sigma_{\rm GF}(v_i,L/a)$ at 
approximately 9 values of the coupling. The very small mistuning can
be corrected by shifting our data for $\Sigma_{\rm GF}$ to the exact target
values by using the relation~\eqref{eq:prop}. These shifted data can be
extrapolated to the continuum in a straightforward
way. Figure~\ref{fig:sigma} shows two types of continuum
extrapolations of the step scaling function
\begin{eqnarray}
  \label{e:fitGF2a}
  \Sigma_{\rm GF}(v_i,L/a) &= \sigma_{\rm GF}(v_i) + \rho_i \times(a/L)^2\,, \\
  \label{e:fitGF2b}
  {1}/{\Sigma_{\rm GF}(v_i,L/a)} &= {1}/{\sigma_{\rm GF}(v_i)}
  + \tilde \rho_i \times(a/L)^2\,.
\end{eqnarray}
Note that the difference between these fit ans\"atze is ${\rm O}(a^4)$. As it
is apparent in Figure~\ref{fig:sigma}, the continuum 
extrapolated values for both fit ans\"atze agree within one standard deviation,
but there is a systematic difference between them. Despite the fact 
that our data shows a very nice $a^2$ scaling, the large cutoff
effects, specially at large values of $u$, induce a systematic effect
in our continuum extrapolations. 
We choose to include an estimate of this systematic uncertainty in the
weights of our fits. 
\begin{equation}
  \label{eq:sys}
  \Delta^{\rm sys} \Sigma_{\rm GF} = 0.05\,\Sigma_{\rm GF}\, 
  \bigg(8\frac{a}{L}\bigg)^4\, \frac{u}{u_{\rm
        max}}\,. 
\end{equation}
This estimate comes from the size of the $a^2$ cutoff effects at
our largest value of the coupling (around 20\% for our coarser
lattice with $L/a=8$), that suggests that the O($a^4$) effects are around
5\%. This effect is added in quadrature to the 
uncertainty in $\Sigma_{\rm GF}$. The result of this procedure is
apparent if one compares Figures~\ref{fig:sigma}
and~\ref{fig:sigma_a4}. The fit functional forms are less constrained
by the coarser lattices, resulting in a better agreement between the
two extrapolation procedures. The price to pay is an increased error
in the extrapolations.  All our final numbers follow this fitting
procedure. The interested 
reader can find a detailed discussion in~\cite{DallaBrida:2016kgh}.  

As stated earlier our non-perturbative data obeys very well the relation
$1/\sigma - 1/u = \text{constant}$ (the 1-loop functional form). This
suggests to fit the continuum 
step scaling function to a functional form
\begin{equation}
  \frac{1}{\sigma_{\rm GF}(v_i)} - \frac{1}{v_i} = Q(v_i)\,,
  \qquad 
  Q(u)= \sum_{k=0}^{n_{\rm sig}-1} c_k u^k\,,
  \label{e:Qfit}
\end{equation}
where the number of fit parameters $n_{\rm sig}$ is varied to check
the consistency of the results. 

Finally one can also combine the continuum extrapolations and the
parametrization of the step scaling functions by fitting
\begin{equation}
  \frac{1}{\Sigma_{\rm GF}(v_i,L/a)} - \frac{1}{v_i} = Q(v_i) +
  \rho(v_i)(a/L)^2\,, 
  \qquad 
  \rho(u)= \sum_{k=0}^{n_\rho-1} \rho_k u^k\,.
  \label{e:Qfit}
\end{equation}
We find good fits when $n_\rho$ is at least 2. Note that this
procedure does not require to shift the data to constant values of
the coupling. All these fitting
procedures produce a remarkable agreement, as is discussed in detail
in~\cite{DallaBrida:2016kgh}.

\subsection{Determination of the $\beta$-function}

\begin{table}[t!]
  \small
  \centering
  \begin{tabular}{ccc|llllll}
    \toprule
    Fit& $n_{\rm sig}$& $n_\rho$& $u_1$&$u_2$&$u_3$& $u_4$ & $s(g_1^2,g_2^2)$\\
  \midrule %\hline
%  obs:   global-fitbeta-straight (\ref{eq:fitw})  
  %  obs:   global-fitbeta-inverse 
  $\Sigma$ &3& --       &  $        5.870(28)$  & $        3.954(22)$  & $        2.976(17)$  & $        2.385(15)$  & $       11.00(20)$ 
  \\ $1/\Sigma$ &1&3    & $        5.843(20)$  & $        3.939(18)$  & $        2.971(16)$  & $        2.385(13)$  & $       10.96(18)$ 
  \\  $1/\Sigma$ &2&3   & $        5.864(26)$  & $        3.944(19)$  & $        2.968(16)$  & $        2.378(14)$  & $       10.90(18)$ 
  \\  $1/\Sigma$ &3&3   & $        5.864(27)$  & $        3.944(21)$  & $        2.968(17)$  & $        2.378(14)$  & $       10.90(19)$ 
  \\ \eqref{eq:global2}, $P$ &2&2 & $        5.872(27)$  & $        3.949(19)$  & $        2.971(16)$  & $        2.379(14)$  & $       10.93(19)$ 
  \\ \eqref{eq:global2}, $P$ &3&3 & $        5.874(28)$  & $        3.951(22)$  & $        2.972(17)$  & $        2.379(14)$  & $       10.93(20)$ 
    \\
  \bottomrule 
  \end{tabular}
  \caption{Coupling sequence \eq{e:recurs} with $u_0=11.31$ 
  and scale factors $s(g_1^2,g_2^2)$ for $g_1^2=2.6723,\,g_2^2=11.31$
  for different fits to cutoff effects and the continuum 
  $\beta$-function. Fits are labeled by $\Sigma$ or $1/\Sigma$ 
  for continuum extrapolations according to \eq{e:fitGF2a},
  \eq{e:fitGF2b} or \eq{eq:global2}. For global fits
  we specify $n_\rho$, while its absence indicates a
  fit of data extrapolated to the continuum at each value of $u=v_i$. 
  }
  \label{tab:ui}
\end{table}
As we have said, our main objective is to compute a ratio of scales,
and in order to do this, we need to determine the $\beta$-function. We
choose the parametrization
\begin{eqnarray}
  \beta(g) = - \frac{g^3}{P(g^2) }\,,\quad P(g^2) = \sum_{k=0}^{n_{\rm
  sig}-1}p_k g^{2k}\,. 
  \label{e:betafct}
\end{eqnarray}
The 1-loop $\beta$-function corresponds to the choice
$n_{\rm sig}=1$. The relation between the $\beta$-function and the
step scaling function $\sigma_{\rm GF}$
\begin{eqnarray}
  \label{e:sigmaint}
  \begin{split}
    \log(2) &= 
    -\int_{\sqrt{u}}^{\sqrt{\sigma(u)}} \frac{\rmd x}{\beta(x)} = 
    \int_{\sqrt{u}}^{\sqrt{\sigma(u)}}\rmd x \frac{P(x^2)
    }{x^3}\\
    &= -\frac{p_0}{2}\left[\frac{1}{\sigma(u)} - \frac{1}{u}\right]
    + \frac{p_1}{2} \log\left[\frac{\sigma(u)}{u}\right]
    + \sum_{k=1}^{n_\mathrm{sig}} \frac{p_{k+1}}{2k} 
    \left[\sigma^{k}(u) - u^{k}\right]\,,
  \end{split}
\end{eqnarray}
is used to fit the coefficients $p_k$. An alternative analysis
consists in combining the continuum extrapolation with the
determination of the $\beta$-function by fitting
\begin{equation}
  \label{eq:global2}
  \log(2) + \widetilde \rho(u)(a/L)^2 = 
  -\int_{\sqrt{u}}^{\sqrt{\Sigma(u,a/L)}} \frac{\rmd x}{\beta(x)}\,.
\end{equation}
Note that this ansatz provides yet another parametrization of the
cutoff effects. 

A quantitative test of the agreement between different functional
forms consists in analyzing the sequence
\begin{equation}
\label{e:recurs}
  u_0=11.31, \quad u_i = \sigma_{\rm GF}^{-1}(u_{i-1})\,, i=1,2,\ldots.
\end{equation}
Table~\ref{tab:ui} contains a sample of the different analysis
considered in~\cite{DallaBrida:2016kgh}. The agreement between
different ans\"atze is remarkable. The last column of table~\ref{tab:ui}
is the scale factor 
\begin{equation}
  \label{eq:scale}
  s(g_1^2,g_2^2) = \exp\left\{
    - \int_{g_1}^{g_2}\frac{dx}{\beta(x)}
  \right\}\,,
\end{equation}
with $g_1^2=2.6723$ and $g_2^2=11.31$. As we will see in the next
section $\gbar^2_{\rm GF}(2L_0)=2.6723(64)$, and therefore this last column is
just the result that we are looking for. 
\begin{figure}[t]
  \centering
  \includegraphics[width=0.9\textwidth]{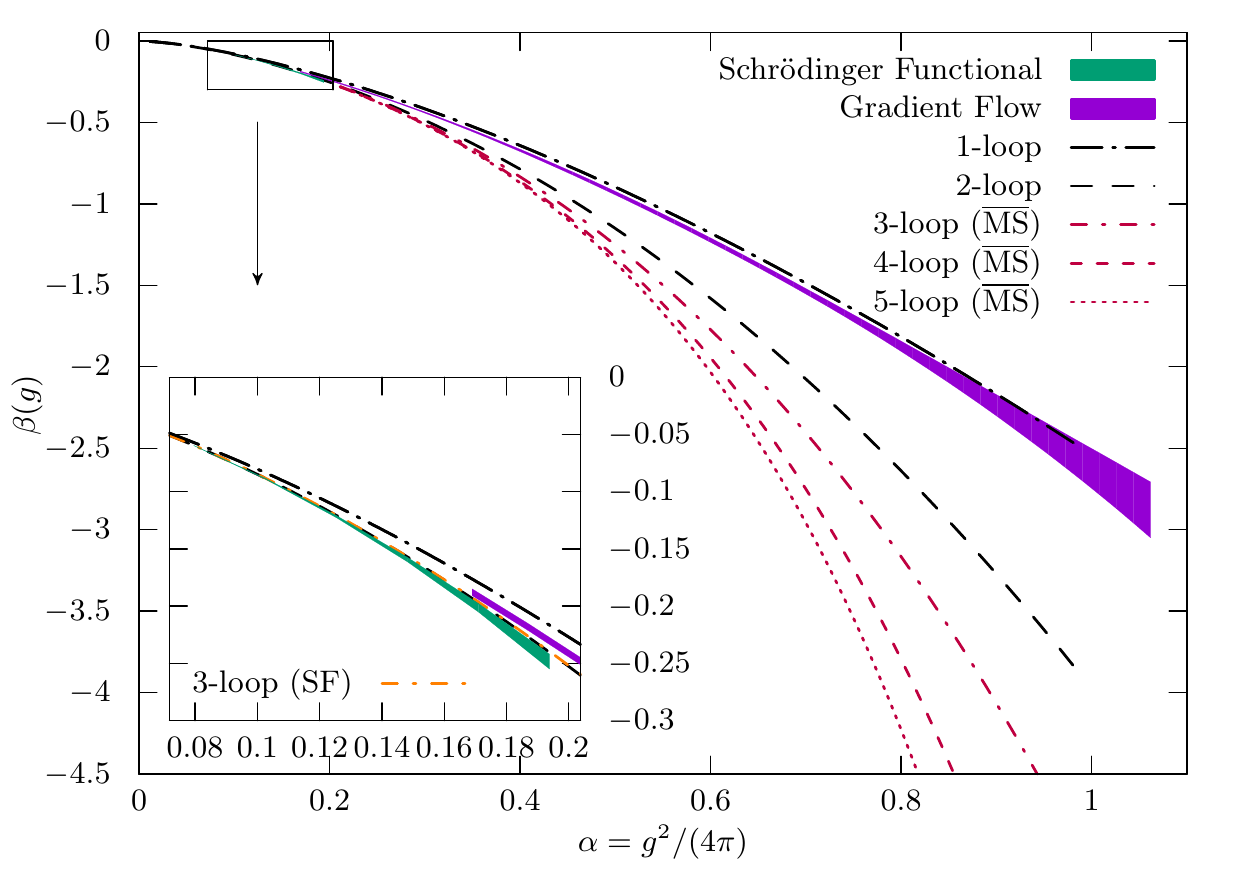}
  \caption{The non-perturbative $\beta$-functions in the SF-scheme from
  \cite{Brida:2016flw} and in the
  GF-scheme~\cite{DallaBrida:2016kgh}. The plotted 1,2-loop universal
  part of the   perturbative expansion 
  can be compared directly, but higher orders of the perturbative series are
  unknown for our finite volume GF-scheme. We give an impression of
  the typical 
  magnitude of higher order perturbative terms in the form of
  the $\overline{\rm MS}$ scheme,
  for which we show curves up to 5-loops. On the other hand the 3-loop
  term is known for the SF scheme.}
  \label{f:beta}
\end{figure}
Figure~\ref{f:beta} shows the results of our non-perturbative running
both in the SF scheme and in the GF scheme. It is remarkable that the
running of the GF coupling follows very closely the 1-loop functional
form but with a slightly different
coefficient. We conclude that
perturbation theory has broken down in most of the range 
that we cover for $\gbar^2_{\rm GF}$. 
The interested reader is invited to read the full
discussion in~\cite{DallaBrida:2016kgh}.

\subsection{The ratio of scales $L_0/\lmax$}

\begin{figure}
  \centering
  \includegraphics[width=0.8\textwidth]{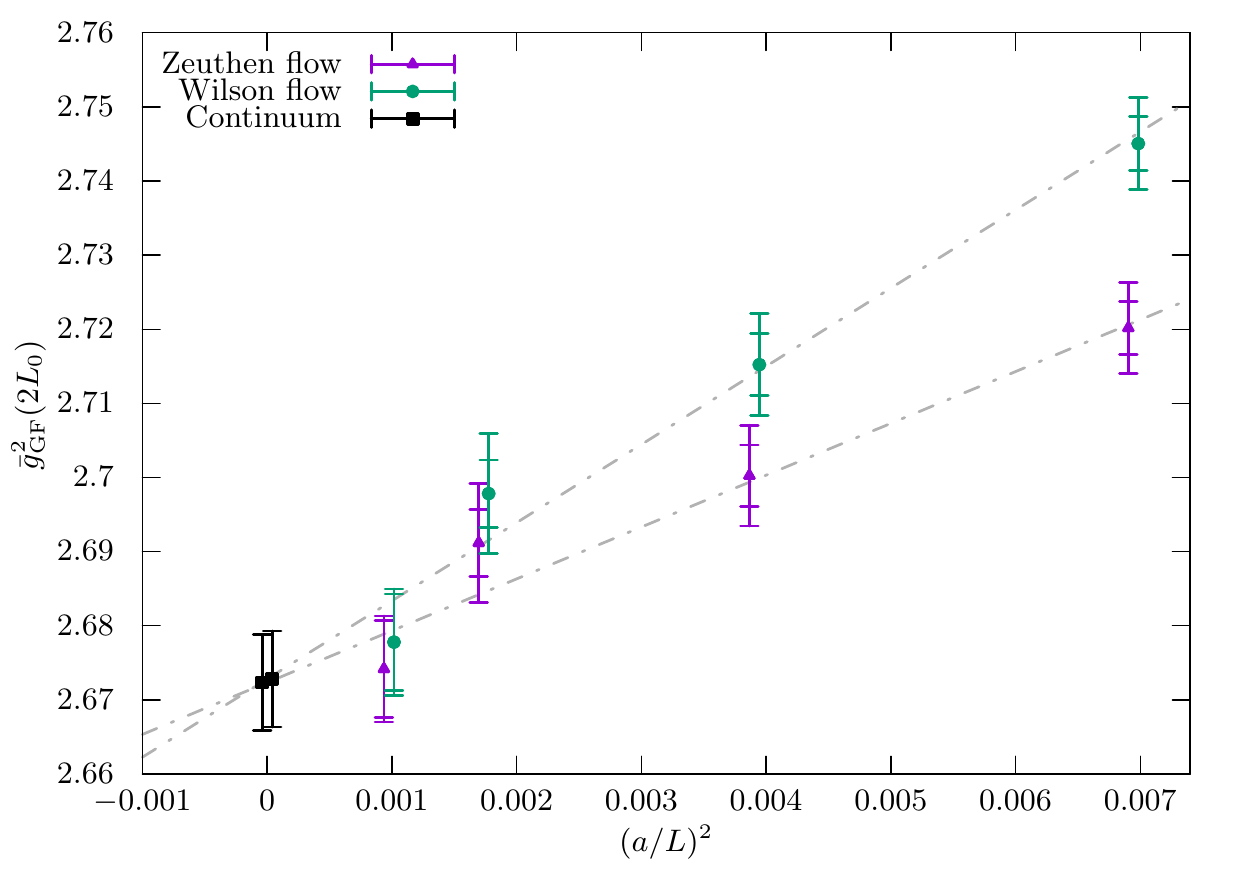}
  \caption{Continuum extrapolation of $\bar \gbar^2_{\rm GF}(2L_0)$
    with the bare parameters determined by the condition
    $\bar g_{\rm SF}^2(L_0) = 2.012$. The continuum
    extrapolation is performed using both the Wilson flow/Clover
    discretization and our preferred setup Zeuthen flow/LW observable
    (the latter shows smaller discretization effects). The two types
    of error bars for each data point correspond to the inclusion or
    not of the propagated error for the SF coupling, cf.~text.}
  \label{fig:Lsw}
\end{figure}

As we have explained the scale $\lmax$ is defined by the
condition $\gbar^2_{\rm GF}(\lmax)=11.31$. On the other hand the
scale $L_0$ is defined by $\gbar^2_{\rm SF}(L_{0})=2.012$. In order to
compute the ratio $\lmax/L_0$ using the $\beta$-function
determined in the previous section, we need to know the value of the
GF coupling at the scale $L_0$. To this end we define the
function
\begin{equation}
  \Phi(u,a/L) = \gbar^2_{\rm GF}(2L)
  \Big|_{\gbar^2_{\rm SF}(L)=u, m(L)=0}\,.
\end{equation}
The procedure is simple, we tune the bare parameters on several
lattices sizes ($L/a=6,8,12,16$) to have $m(L)=0$ and $\gbar^2_{\rm
  SF}(L)=2.012$. Note that here we use the Wilson gauge action
  for the reason explained in \sect{s:sint}. Then
we compute the value of the GF coupling at the 
same values of the bare parameters, but on lattices twice as large. 
The error in the tuning of the SF coupling is propagated to the GF coupling by
using leading order perturbation theory. This
procedure has several advantages. On the one hand the SF coupling is
computed only on relatively small lattices. This is convenient, since
the SF coupling requires large statistics, but its cutoff effects
are very small. On the other hand the GF coupling is computed on
larger lattices, to avoid large cutoff effects, but with
relatively modest statistics one can achieve high statistical
precision. 

Recall that the SF coupling is defined with a
background field, while the boundary conditions of our gradient
flow scheme correspond to a zero background field. The connection
between the couplings is established via the common bare parameters
defined by the condition $\gbar^2_{\rm SF}(L)=u, m(L)=0$, together
with the resolution $a/L$.

Once the estimates of $\Phi(2.012,a/L)$ are determined, we can take the
continuum limit and obtain
\begin{equation}
  \gbar^2_{\rm GF}(2L_0)  =  \lim_{a/L\rightarrow 0}
  \Phi(2.012,a/L)  = 2.6723(64)\,.
\end{equation}
Figure~\ref{fig:Lsw} shows the continuum extrapolation with two
different choices of lattice flow equations.

Given the $\beta$-function determined in the previous
section, together with this result, we now obtain (see
equation~\eqref{eq:scale}) 
\begin{equation}
  \frac{\lmax}{L_0} = 2\times s(2.6723(64), 11.31) = 21.86(42)\,.
\end{equation}
For the value $ s(2.6723, 11.31) $ we choose the last row of
table~\ref{tab:ui} which has the largest error of all the considered
analyses. As indicated, our result for the scale factor also contains the error in
$\gbar^2_{\rm GF}(2L_0)$ from the matching procedure.

%%% Local Variables:
%%% mode: latex
%%% TeX-master: "lat16-proceedings"
%%% End:

%% file: acknow.tex
\begin{center}
{\bf Acknowledgments}
\end{center}
This work was done as part of the ALPHA collaboration research programme. 
We thank our colleagues in the ALPHA collaboration,
in particular C.~Pena and U.~Wolff for many useful discussions.

This project has benefited from the joint production of gauge field
ensembles with a project computing the running of quark masses. We
thank I. Campos, C. Pena and D. Preti for this collaboration. We also
thank Pol Vilaseca who computed the used one-loop coefficient $\tilde
c_{\rm t}$.  

We thank the computer centres at HLRN (bep00040) and NIC at DESY,
Zeuthen for providing computing resources and support. We are indebted
to Isabel Campos and thank her and the staff at the University of
Cantabria at IFCA in the Altamira HPC facility for computer resources
and technical support.

S.~Sint gratefully acknowledges support by SFI under grant 11/RFP/PHY3218.
P.F.~acknowledges financial support from the Spanish MINECO's ``Centro de 
Excelencia Severo Ochoa'' Programme under grant SEV-2012-0249, as well
as from the MCINN grant FPA2012-31686. M.D.B. thanks the Theoretical
Physics Department at CERN for the hospitality and support.

%% file: lat16-proceedings.bbl
\providecommand{\href}[2]{#2}\begingroup\raggedright\begin{thebibliography}{10}

\bibitem{Brida:2016flw}
M.~Dalla~Brida, P.~Fritzsch, T.~Korzec, A.~Ramos, S.~Sint and R.~Sommer,
  \emph{{Determination of the QCD $\Lambda$-parameter and the accuracy of
  perturbation theory at high energies}},
  \href{http://dx.doi.org/10.1103/PhysRevLett.117.182001}{\emph{Phys. Rev.
  Lett.} {\bf 117} (2016) 182001}, [\href{http://arxiv.org/abs/1604.06193}{{\tt
  1604.06193}}].

\bibitem{DallaBrida:2016kgh}
{\scshape ALPHA} collaboration, M.~Dalla~Brida, P.~Fritzsch, T.~Korzec,
  A.~Ramos, S.~Sint and R.~Sommer, \emph{{Slow running of the Gradient Flow
  coupling from 200 MeV to 4 GeV in $N_{\rm f}=3$ QCD}}, {\emph{{{}}} (2016) },
  [\href{http://arxiv.org/abs/1607.06423}{{\tt 1607.06423}}].

\bibitem{Bruno:2014jqa}
M.~Bruno et~al., \emph{{Simulation of QCD with $N_{f} = 2 + 1$ flavors of
  non-perturbatively improved Wilson fermions}},
  \href{http://dx.doi.org/10.1007/JHEP02(2015)043}{\emph{JHEP} {\bf 02} (2015)
  043}, [\href{http://arxiv.org/abs/1411.3982}{{\tt 1411.3982}}].

\bibitem{Bruno:2016plf}
M.~Bruno, T.~Korzec and S.~Schaefer, \emph{{Setting the scale for the CLS $2 +
  1$ flavor ensembles}}, {\emph{{{}}} ({2016}) },
  [\href{http://arxiv.org/abs/{1608.08900}}{{\tt {1608.08900}}}].

\bibitem{Brida:2014joa}
P.~Fritzsch, M.~Dalla~Brida, T.~Korzec, A.~Ramos, S.~Sint and R.~Sommer,
  \emph{{Towards a new determination of the QCD Lambda parameter from running
  couplings in the three-flavour theory}}, {\emph{PoS} {\bf LATTICE2014} (2014)
  291}, [\href{http://arxiv.org/abs/1411.7648}{{\tt 1411.7648}}].

\bibitem{Brida:2015gqj}
M.~Dalla~Brida, P.~Fritzsch, T.~Korzec, A.~Ramos, S.~Sint and R.~Sommer,
  \emph{{A status update on the determination of ${\Lambda}_{\overline{\rm
  MS}}^{N_{\rm f}=3}$ by the ALPHA collaboration}},  in \emph{{Proceedings,
  33rd International Symposium on Lattice Field Theory (Lattice 2015)}}, 2015.
\newblock \href{http://arxiv.org/abs/1511.05831}{{\tt 1511.05831}}.

\bibitem{Luscher:1998pe}
M.~{L\"uscher}, \emph{{Advanced lattice QCD}},  in \emph{{Probing the standard
  model of particle interactions. Proceedings, Summer School in Theoretical
  Physics, NATO Advanced Study Institute, 68th session, Les Houches, France,
  July 28-September 5, 1997. Pt. 1, 2}}, pp.~229--280, 1998.
\newblock \href{http://arxiv.org/abs/hep-lat/9802029}{{\tt hep-lat/9802029}}.

\bibitem{Sommer:2006sj}
R.~Sommer, \emph{{Non-perturbative QCD: Renormalization, O(a)-improvement and
  matching to Heavy Quark Effective Theory}},  in \emph{{Workshop on
  Perspectives in Lattice QCD Nara, Japan, October 31-November 11, 2005}},
  2006.
\newblock \href{http://arxiv.org/abs/hep-lat/0611020}{{\tt hep-lat/0611020}}.

\bibitem{Bernreuther:1981sg}
W.~Bernreuther and W.~Wetzel, \emph{{Decoupling of Heavy Quarks in the Minimal
  Subtraction Scheme}}, \href{http://dx.doi.org/10.1016/0550-3213(82)90288-7,
  10.1016/S0550-3213(97)00811-0}{\emph{Nucl. Phys.} {\bf B197} (1982)
  228--236}.

\bibitem{Chetyrkin:2005ia}
K.~G. Chetyrkin, J.~H. K\"uhn and C.~Sturm, \emph{{QCD decoupling at four
  loops}}, \href{http://dx.doi.org/10.1016/j.nuclphysb.2006.03.020}{\emph{Nucl.
  Phys.} {\bf B744} (2006) 121--135},
  [\href{http://arxiv.org/abs/hep-ph/0512060}{{\tt hep-ph/0512060}}].

\bibitem{brida2016}
M.~Dalla~Brida, T.~Korzec, S.~Sint and P.~Vilaseca, \emph{{High precision
  renormalization of the non-singlet axial current in lattice QCD with Wilson
  quarks}}, {\emph{{\emph{in preparation}}} }.

\bibitem{Brida:2016rmy}
M.~Dalla~Brida, S.~Sint and P.~Vilaseca, \emph{{The chirally rotated
  Schr\"odinger functional: theoretical expectations and perturbative tests}},
  \href{http://dx.doi.org/10.1007/JHEP08(2016)102}{\emph{JHEP} {\bf 08} (2016)
  102}, [\href{http://arxiv.org/abs/1603.00046}{{\tt 1603.00046}}].

\bibitem{SFcoupinpreparation}
M.~Dalla~Brida, P.~Fritzsch, T.~Korzec, A.~Ramos, S.~Sint and R.~Sommer. in
  preparation.

\bibitem{Bruno:2016gvs}
M.~Bruno, M.~Dalla~Brida, P.~Fritzsch, T.~Korzec, A.~Ramos, S.~Schaefer et~al.,
  \emph{{The determination of $\alpha_s$ by the ALPHA collaboration}},  in
  \emph{{6th Workshop on Theory, Phenomenology and Experiments in Flavour
  Physics: Interplay of Flavour Physics with electroweak symmetry breaking
  (Capri 2016) Anacapri, Capri, Italy, June 11, 2016}}, 2016.
\newblock \href{http://arxiv.org/abs/1611.05750}{{\tt 1611.05750}}.

\bibitem{Luscher:1992an}
M.~{L\"uscher}, R.~Narayanan, P.~Weisz and U.~Wolff, \emph{{The Schr\"odinger
  functional: a renormalizable probe for non-Abelian gauge theories}},
  \href{http://dx.doi.org/10.1016/0550-3213(92)90466-O}{\emph{Nucl. Phys.} {\bf
  B384} (1992) 168--228}, [\href{http://arxiv.org/abs/hep-lat/9207009}{{\tt
  hep-lat/9207009}}].

\bibitem{Sint:1993un}
S.~Sint, \emph{{On the Schr\"odinger functional in QCD}},
  \href{http://dx.doi.org/10.1016/0550-3213(94)90228-3}{\emph{Nucl. Phys.} {\bf
  B421} (1994) 135--158}, [\href{http://arxiv.org/abs/hep-lat/9312079}{{\tt
  hep-lat/9312079}}].

\bibitem{Luscher:1993gh}
M.~{L\"uscher}, R.~Sommer, P.~Weisz and U.~Wolff, \emph{{A Precise
  determination of the running coupling in the SU(3) Yang-Mills theory}},
  \href{http://dx.doi.org/10.1016/0550-3213(94)90629-7}{\emph{Nucl. Phys.} {\bf
  B413} (1994) 481--502}, [\href{http://arxiv.org/abs/hep-lat/9309005}{{\tt
  hep-lat/9309005}}].

\bibitem{Sint:2012ae}
S.~Sint and P.~Vilaseca, \emph{{Lattice artefacts in the Schr\"odinger
  Functional coupling for strongly interacting theories}}, {\emph{PoS} {\bf
  LATTICE2012} (2012) 031}, [\href{http://arxiv.org/abs/1211.0411}{{\tt
  1211.0411}}].

\bibitem{Bode:1999sm}
{\scshape ALPHA} collaboration, A.~Bode, P.~Weisz and U.~Wolff, \emph{{Two loop
  computation of the Schr\"odinger functional in lattice QCD}},
  \href{http://dx.doi.org/10.1016/S0550-3213(00)00187-5}{\emph{Nucl. Phys.}
  {\bf B576} (2000) 517--539},
  [\href{http://arxiv.org/abs/hep-lat/9911018}{{\tt hep-lat/9911018}}].

\bibitem{Luscher:1995nr}
M.~{L\"uscher} and P.~Weisz, \emph{{Two loop relation between the bare lattice
  coupling and the MS coupling in pure SU(N) gauge theories}},
  \href{http://dx.doi.org/10.1016/0370-2693(95)00250-O}{\emph{Phys. Lett.} {\bf
  B349} (1995) 165--169}, [\href{http://arxiv.org/abs/hep-lat/9502001}{{\tt
  hep-lat/9502001}}].

\bibitem{Christou:1998wk}
C.~Christou, H.~Panagopoulos, A.~Feo and E.~Vicari, \emph{{The two loop
  relation between the bare lattice coupling and the MS-bar coupling in QCD
  with Wilson fermions}},
  \href{http://dx.doi.org/10.1016/S0370-2693(98)00278-0}{\emph{Phys. Lett.}
  {\bf B426} (1998) 121--124}.

\bibitem{Christou:1998ws}
C.~Christou, A.~Feo, H.~Panagopoulos and E.~Vicari, \emph{{The three loop
  $\beta$-function of $SU(N)$ lattice gauge theories with Wilson fermions}},
  \href{http://dx.doi.org/10.1016/S0550-3213(01)00268-1,
  10.1016/S0550-3213(98)00248-X}{\emph{Nucl. Phys.} {\bf B525} (1998)
  387--400}, [\href{http://arxiv.org/abs/hep-lat/9801007}{{\tt
  hep-lat/9801007}}].

\bibitem{MS:4loop1}
T.~van Ritbergen, J.~A.~M. Vermaseren and S.~A. Larin, \emph{The four loop beta
  function in quantum chromodynamics}, {\emph{Phys. Lett.} {\bf B400} (1997)
  379--384}, [\href{http://arxiv.org/abs/hep-ph/9701390}{{\tt
  hep-ph/9701390}}].

\bibitem{Czakon:2004bu}
M.~Czakon, \emph{{The Four-loop QCD beta-function and anomalous dimensions}},
  \href{http://dx.doi.org/10.1016/j.nuclphysb.2005.01.012}{\emph{Nucl.Phys.}
  {\bf B710} (2005) 485--498}, [\href{http://arxiv.org/abs/hep-ph/0411261}{{\tt
  hep-ph/0411261}}].

\bibitem{Yamada:2004ja}
{\scshape JLQCD, CP-PACS} collaboration, N.~Yamada et~al.,
  \emph{{Non-perturbative O(a)-improvement of Wilson quark action in
  three-flavor QCD with plaquette gauge action}},
  \href{http://dx.doi.org/10.1103/PhysRevD.71.054505}{\emph{Phys.Rev.} {\bf
  D71} (2005) 054505}, [\href{http://arxiv.org/abs/hep-lat/0406028}{{\tt
  hep-lat/0406028}}].

\bibitem{Nf3tuning}
P.~Fritzsch and T.~Korzec, \emph{{Simulating the QCD Schr\"odinger Functional
  with three massless quark flavors}}, {\emph{in preparation} (2016) }.

\bibitem{Luscher:2012av}
M.~{L\"uscher} and S.~Schaefer, \emph{{Lattice QCD with open boundary
  conditions and twisted-mass reweighting}},
  \href{http://dx.doi.org/10.1016/j.cpc.2012.10.003}{\emph{Comput. Phys.
  Commun.} {\bf 184} (2013) 519--528},
  [\href{http://arxiv.org/abs/1206.2809}{{\tt 1206.2809}}].

\bibitem{Aoki:2009tf}
{\scshape PACS-CS} collaboration, S.~Aoki et~al., \emph{{Precise determination
  of the strong coupling constant in $N_f = 2+1$ lattice QCD with the
  Schr\"odinger functional scheme}},
  \href{http://dx.doi.org/10.1088/1126-6708/2009/10/053}{\emph{JHEP} {\bf 0910}
  (2009) 053}, [\href{http://arxiv.org/abs/0906.3906}{{\tt 0906.3906}}].

\bibitem{Luscher:1996vw}
M.~{L\"uscher} and P.~Weisz, \emph{{O(a) improvement of the axial current in
  lattice QCD to one loop order of perturbation theory}},
  \href{http://dx.doi.org/10.1016/0550-3213(96)00448-8}{\emph{Nucl. Phys.} {\bf
  B479} (1996) 429--458}, [\href{http://arxiv.org/abs/hep-lat/9606016}{{\tt
  hep-lat/9606016}}].

\bibitem{deDivitiis:1994yz}
{\scshape ALPHA} collaboration, G.~de~Divitiis, R.~Frezzotti, M.~Guagnelli,
  M.~{L\"uscher}, R.~Petronzio, R.~Sommer et~al., \emph{{Universality and the
  approach to the continuum limit in lattice gauge theory}},
  \href{http://dx.doi.org/10.1016/0550-3213(94)00019-B}{\emph{Nucl. Phys.} {\bf
  B437} (1995) 447--470}, [\href{http://arxiv.org/abs/hep-lat/9411017}{{\tt
  hep-lat/9411017}}].

\bibitem{Ramos:2015dla}
A.~Ramos, \emph{{The Yang-Mills gradient flow and renormalization}},
  {\emph{PoS} {\bf LATTICE2014} (2015) 017},
  [\href{http://arxiv.org/abs/1506.00118}{{\tt 1506.00118}}].

\bibitem{Fritzsch:2013je}
P.~Fritzsch and A.~Ramos, \emph{{The gradient flow coupling in the Schrödinger
  Functional}}, \href{http://dx.doi.org/10.1007/JHEP10(2013)008}{\emph{JHEP}
  {\bf 10} (2013) 008}, [\href{http://arxiv.org/abs/1301.4388}{{\tt
  1301.4388}}].

\bibitem{Narayanan:2006rf}
R.~Narayanan and H.~Neuberger, \emph{{Infinite N phase transitions in continuum
  Wilson loop operators}},
  \href{http://dx.doi.org/10.1088/1126-6708/2006/03/064}{\emph{JHEP} {\bf 0603}
  (2006) 064}, [\href{http://arxiv.org/abs/hep-th/0601210}{{\tt
  hep-th/0601210}}].

\bibitem{Luscher:2010iy}
M.~L{\"u}scher, \emph{{Properties and uses of the Wilson flow in lattice QCD}},
  \href{http://dx.doi.org/10.1007/JHEP08(2010)071}{\emph{JHEP} {\bf 1008}
  (2010) 071}, [\href{http://arxiv.org/abs/1006.4518}{{\tt 1006.4518}}].

\bibitem{Luscher:2011bx}
M.~L{\"u}scher and P.~Weisz, \emph{{Perturbative analysis of the gradient flow
  in non-abelian gauge theories}},
  \href{http://dx.doi.org/10.1007/JHEP02(2011)051}{\emph{JHEP} {\bf 1102}
  (2011) 051}, [\href{http://arxiv.org/abs/1101.0963}{{\tt 1101.0963}}].

\bibitem{DelDebbio:2004xh}
L.~Del~Debbio, G.~M. Manca and E.~Vicari, \emph{{Critical slowing down of
  topological modes}},
  \href{http://dx.doi.org/10.1016/j.physletb.2004.05.038}{\emph{Phys. Lett.}
  {\bf B594} (2004) 315}, [\href{http://arxiv.org/abs/hep-lat/0403001}{{\tt
  hep-lat/0403001}}].

\bibitem{Fritzsch:2013yxa}
P.~Fritzsch, A.~Ramos and F.~Stollenwerk, \emph{{Critical slowing down and the
  gradient flow coupling in the Schr\"odinger functional}}, {\emph{PoS} {\bf
  Lattice2013} (2013) 461}, [\href{http://arxiv.org/abs/1311.7304}{{\tt
  1311.7304}}].

\bibitem{Aoki:1998qd}
S.~Aoki, R.~Frezzotti and P.~Weisz, \emph{{Computation of the improvement
  coefficient $c_{\rm SW}$ to one loop with improved gluon actions}},
  \href{http://dx.doi.org/10.1016/S0550-3213(98)00742-1}{\emph{Nucl. Phys.}
  {\bf B540} (1999) 501}, [\href{http://arxiv.org/abs/hep-lat/9808007}{{\tt
  hep-lat/9808007}}].

\bibitem{Takeda:2003he}
S.~Takeda, S.~Aoki and K.~Ide, \emph{{A Perturbative determination of $O(a)$
  boundary improvement coefficients for the Schr\"odinger functional coupling
  at one loop with improved gauge actions}},
  \href{http://dx.doi.org/10.1103/PhysRevD.68.014505}{\emph{Phys. Rev.} {\bf
  D68} (2003) 014505}, [\href{http://arxiv.org/abs/hep-lat/0304013}{{\tt
  hep-lat/0304013}}].

\bibitem{pol}
P.~Vilaseca{\emph{{\rm ,} private communication} (2015) }.

\bibitem{Ramos:2015baa}
A.~Ramos and S.~Sint, \emph{{Symanzik improvement of the gradient flow in
  lattice gauge theories}},
  \href{http://dx.doi.org/10.1140/epjc/s10052-015-3831-9}{\emph{Eur. Phys. J.}
  {\bf C76} (2016) 15}, [\href{http://arxiv.org/abs/1508.05552}{{\tt
  1508.05552}}].

\bibitem{Sommer:2014mea}
R.~Sommer, \emph{{Scale setting in lattice QCD}}, {\emph{PoS} {\bf LATTICE2013}
  (2014) 015}, [\href{http://arxiv.org/abs/1401.3270}{{\tt 1401.3270}}].

\bibitem{Aoki:2016frl}
S.~Aoki et~al., \emph{{Review of lattice results concerning low-energy particle
  physics}}, {\emph{{{}}} (2016) },
  [\href{http://arxiv.org/abs/1607.00299}{{\tt 1607.00299}}].

\bibitem{Bietenholz:2010jr}
W.~Bietenholz et~al., \emph{{Tuning the strange quark mass in lattice
  simulations}},
  \href{http://dx.doi.org/10.1016/j.physletb.2010.05.067}{\emph{Phys. Lett.}
  {\bf B690} (2010) 436--441}, [\href{http://arxiv.org/abs/1003.1114}{{\tt
  1003.1114}}].

\bibitem{Gasser:1984gg}
J.~Gasser and H.~Leutwyler, \emph{{Chiral Perturbation Theory: Expansions in
  the Mass of the Strange Quark}},
  \href{http://dx.doi.org/10.1016/0550-3213(85)90492-4}{\emph{Nucl. Phys.} {\bf
  B250} (1985) 465--516}.

\bibitem{Sint:1995ch}
S.~Sint and R.~Sommer, \emph{{The Running coupling from the QCD Schr\"odinger
  functional: A One loop analysis}},
  \href{http://dx.doi.org/10.1016/0550-3213(96)00020-X}{\emph{Nucl. Phys.} {\bf
  B465} (1996) 71--98}, [\href{http://arxiv.org/abs/hep-lat/9508012}{{\tt
  hep-lat/9508012}}].

\bibitem{Luscher:1996sc}
M.~{L\"uscher}, S.~Sint, R.~Sommer and P.~Weisz, \emph{{Chiral symmetry and
  O(a) improvement in lattice QCD}},
  \href{http://dx.doi.org/10.1016/0550-3213(96)00378-1}{\emph{Nucl. Phys.} {\bf
  B478} (1996) 365--400}, [\href{http://arxiv.org/abs/hep-lat/9605038}{{\tt
  hep-lat/9605038}}].

\bibitem{Luscher:2011kk}
M.~L{\"u}scher and S.~Schaefer, \emph{{Lattice QCD without topology barriers}},
  \href{http://dx.doi.org/10.1007/JHEP07(2011)036}{\emph{JHEP} {\bf 1107}
  (2011) 036}, [\href{http://arxiv.org/abs/1105.4749}{{\tt 1105.4749}}].

\bibitem{Bruno:2014ufa}
{\scshape ALPHA} collaboration, M.~Bruno, J.~Finkenrath, F.~Knechtli, B.~Leder
  and R.~Sommer, \emph{{Effects of Heavy Sea Quarks at Low Energies}},
  \href{http://dx.doi.org/10.1103/PhysRevLett.114.102001}{\emph{Phys. Rev.
  Lett.} {\bf 114} (2015) 102001}, [\href{http://arxiv.org/abs/1410.8374}{{\tt
  1410.8374}}].

\bibitem{Weinberg:1980wa}
S.~Weinberg, \emph{{Effective gauge theories}},
  \href{http://dx.doi.org/10.1016/0370-2693(80)90660-7}{\emph{Phys.Lett.} {\bf
  B91} (1980) 51}.

\bibitem{Schroder:2005hy}
Y.~{Schr{\"o}der} and M.~Steinhauser, \emph{{Four-loop decoupling relations for
  the strong coupling}},
  \href{http://dx.doi.org/10.1088/1126-6708/2006/01/051}{\emph{JHEP} {\bf 01}
  (2006) 051}, [\href{http://arxiv.org/abs/hep-ph/0512058}{{\tt
  hep-ph/0512058}}].

\bibitem{Agashe:2014kda}
{\scshape Particle Data Group} collaboration, K.~A. Olive et~al., \emph{{Review
  of Particle Physics}},
  \href{http://dx.doi.org/10.1088/1674-1137/38/9/090001}{\emph{Chin. Phys.}
  {\bf C38} (2014) 090001}.

\bibitem{Baikov:2016tgj}
P.~A. Baikov, K.~G. Chetyrkin and J.~H. K\"uhn, \emph{{Five-Loop Running of the
  QCD coupling constant}}, {\emph{{{}}} (2016) },
  [\href{http://arxiv.org/abs/1606.08659}{{\tt 1606.08659}}].

\bibitem{Luthe:2016ima}
T.~Luthe, A.~Maier, P.~Marquard and Y.~{Schr\"oder}, \emph{{Towards the
  five-loop Beta function for a general gauge group}},
  \href{http://dx.doi.org/10.1007/JHEP07(2016)127}{\emph{JHEP} {\bf 07} (2016)
  127}, [\href{http://arxiv.org/abs/1606.08662}{{\tt 1606.08662}}].

\end{thebibliography}\endgroup
